\documentclass[12pt]{article}
\usepackage{graphicx}
\usepackage{epsfig}
\usepackage{epstopdf}
\DeclareGraphicsExtensions{.pdf,.eps,.png,.jpg,.mps}

\usepackage{cite}

\def\lsim{\mathrel{\rlap{\lower3pt\hbox{\hskip0pt$\sim$}}
     \raise1pt\hbox{$<$}}}         
\def\gsim{\mathrel{\rlap{\lower4pt\hbox{\hskip1pt$\sim$}}
     \raise1pt\hbox{$>$}}}         

\usepackage{amsmath}
\usepackage{amsfonts}

\begin{document}
\begin{titlepage}

\centerline{\Large \bf Multifactor Risk Models and Heterotic CAPM}
\medskip

\centerline{Zura Kakushadze$^\S$$^\dag$\footnote{\, Zura Kakushadze, Ph.D., is the President of Quantigic$^\circledR$ Solutions LLC,
and a Full Professor at Free University of Tbilisi. Email: zura@quantigic.com} and Willie Yu$^\sharp$\footnote{\, Willie Yu, Ph.D., is a Research Fellow at Duke-NUS Medical School. Email: willie.yu@duke-nus.edu.sg}}
\bigskip

\centerline{\em $^\S$ Quantigic$^\circledR$ Solutions LLC}
\centerline{\em 1127 High Ridge Road \#135, Stamford, CT 06905\,\,\footnote{\, DISCLAIMER: This address is used by the corresponding author for no
purpose other than to indicate his professional affiliation as is customary in
publications. In particular, the contents of this paper
are not intended as an investment, legal, tax or any other such advice,
and in no way represent views of Quantigic$^\circledR$ Solutions LLC,
the website \underline{www.quantigic.com} or any of their other affiliates.
}}
\centerline{\em $^\dag$ Free University of Tbilisi, Business School \& School of Physics}
\centerline{\em 240, David Agmashenebeli Alley, Tbilisi, 0159, Georgia}
\centerline{\em $^\sharp$ Centre for Computational Biology, Duke-NUS Medical School}
\centerline{\em 8 College Road, Singapore 169857}
\medskip
\centerline{(January 24, 2016)}

\bigskip
\medskip
\centerline{\it ZK: To my children Mirabelle and Maximilien Kakushadze}
\medskip
\centerline{\it WY: To my parents Albert and Ribena Yu}
\bigskip
\medskip

\begin{abstract}
{}We give a complete algorithm and source code for constructing general multifactor risk models (for equities) via any combination of style factors, principal components (betas) and/or industry factors. For short horizons we employ the Russian-doll risk model construction to obtain a nonsingular factor covariance matrix. This generalizes the heterotic risk model construction to include arbitrary non-industry risk factors as well as industry risk factors with generic ``weights". The aim of sharing our proprietary know-how with the investment community is to encourage organic risk model building. The presentation is intended to be essentially self-contained and pedagogical. So, stop wasting money and complaining, start building risk models and enjoy!\footnote{\, This is the last paper in the trilogy, which contains ``Russian-Doll Risk Models" \cite{RusDoll} and ``Heterotic Risk Models" \cite{Het}.}
\end{abstract}
\medskip
\end{titlepage}

\newpage

\section{Introduction}

{}Quantitative traders build their own alphas and trading strategies. However, only a small fraction of quantitative traders build their own risk models. How come? Especially considering that it is precisely the risk models -- one way or another -- that turn alphas (or expected returns, trading signals, etc.) into portfolio holdings.

{}A typical answer is ``we focus on alphas",\footnote{\, To traders, alphas make money. Risk models just help make money. Hence the priorities.} ``we do not have the required data", ``we do not have time/human resources", etc. So, most quantitative traders end up buying off-the-shelf standardized commercial risk models with at least some understanding that what they are buying is not built for their kind of trading or tested on real-life alphas.\footnote{\, The majority of clients of the standardized commercial risk model providers are large institutionals such as mutual and pension funds that have no need for customization but require standardization for inter- and intra-institutional risk reporting, etc. \cite{CustomRM}.} In fact, many quant traders even complain about this.

{}The prosaic answer to the above question is actually much simpler. Most quant traders do not know how to build high quality risk models from scratch. It takes time and experience to develop expertise for such model building and most quant traders indeed spend much of their time developing alphas. There are some quant shops that develop their own risk models, but they appear to be in the minority.

{}The purpose of this paper is to change the aforesaid landscape by giving a complete algorithm and source code for constructing general multifactor risk models\footnote{\,
For a partial list of works related to factor risk models, see
\cite{Q1}, \cite{Q2}, \cite{Q3}, \cite{Q4}, \cite{Q5}, \cite{Q6}, \cite{Q7}, \cite{Q8}, \cite{Q9}, \cite{Q10},
\cite{Q11}, \cite{Q12}, \cite{Q13}, \cite{Q14}, \cite{Q15}, \cite{Q16}, \cite{Q17}, \cite{Q18}, \cite{Q19}, \cite{Q20},
\cite{Q21}, \cite{Q22}, \cite{Q23}, \cite{Q24}, \cite{Q26, Q25}, \cite{Q27}, \cite{Q28}, \cite{Q29, Q30},
\cite{Q31, Q32, Q33}, \cite{Q34}, \cite{Q35}, \cite{Q36}, \cite{Q37, Q38, Q39, Q40},
\cite{Q41}, \cite{Q42, Q43}, \cite{Q44, Q45}, \cite{Q46}, \cite{Q47}, \cite{Q48}, \cite{Q49}, \cite{Q50},
\cite{Q51}, \cite{Q52}, \cite{Q53}, \cite{Q54}, \cite{Q55, Q56}, (Kakushadze 2015a,b,c,d), \cite{CustomRM},
\cite{Q62}, \cite{Q63}, \cite{Q64}, \cite{Q65}, \cite{Q66}, \cite{Q67}, \cite{Q68}, \cite{Q69}, \cite{Q70},
\cite{Q71}, \cite{Q72}, \cite{Q73}, \cite{Q74, Q75}, \cite{Q76}, \cite{Q77}, \cite{Q78}, \cite{Q79}, \cite{Q80},
\cite{Q81}, \cite{Q82}, \cite{Q83}, \cite{Q84}, \cite{Q85, Q86, Q87}, \cite{Q88}, \cite{Q89}, \cite{Q90, Q91}, \cite{Q92},
\cite{Q93, Q94}, \cite{Q95, Q96}, \cite{Q97}, \cite{Q98}, \cite{Q99}, \cite{Q100}, \cite{Q101}, \cite{Q102},
and references therein.
}
for equities\footnote{\, Albeit the general methods we discuss here can be applied to other instruments as well.} via any combination of style factors, principal components and/or industry factors. For short horizons we employ the Russian-doll risk model construction \cite{RusDoll} to obtain a nonsingular factor covariance matrix. This generalizes the heterotic risk model construction \cite{Het} to include arbitrary non-industry as well as industry risk factors with generic ``weights". By sharing our proprietary know-how we hope to encourage organic risk model building.

{}The main difficulty in dealing with generic risk factors is that using the standard ``lore" for computing the factor covariance matrix and specific (idiosyncratic) risk, which is done via a (weighted) regression, yields total variances for stocks which do not agree with sample variances. The exception is when the risk factors are based on principal components of the sample covariance (or correlation) matrix or its sub-matrices corresponding to ``clusters" such as sub-industries as in the heterotic risk model construction. In this case the total risks automatically work out. However, if we include, e.g., style factors, this method produces incorrect total variances, which is often overlooked in practice. Our algorithm rectifies this shortcoming by design.

{}In Section \ref{sec.2} we review multifactor models, the requirement that a factor model reproduce sample variances, and why it is (not) satisfied for principal component (general) risk factors. In Section \ref{sec.3} we discuss industry risk factors and the heterotic risk model construction. In Section \ref{sec.4} we give our algorithm for dealing with general risk factors. However, the paper does not end there. This is because for short historical lookbacks, which is the case in short-horizon models, the sample factor covariance matrix is singular and itself needs to be modeled. This, in turn, is due to the fact that the number of industry factors is too large (in three digits). In Section \ref{sec.5} we apply the Russian-doll construction to general factor models and give a complete embedding resulting in a nonsingular factor covariance matrix and consequently, an invertible multifactor model covariance matrix. We also explain why adding a few style or other non-industry risk factors (e.g., principal components) to ubiquitous industry risk factors adds little to no value.\footnote{\, And this is how style factors are traditionally treated including in standardized commercial risk models. These offerings aver that style factors do add value. However, there are a few caveats there. First, in such models the industry factors are not taken at the most granular level and consequently are not as numerous, typically, around 60 or fewer, so augmenting them with 10 or so style factors a priori may add some value. Second, these models are not necessarily tested on real-life trading alphas. Third, they are geared toward longer horizons, i.e., lower turnover strategies with lower Sharpe ratios. As we discuss in Section \ref{sec.5}, adding industry classification granularity increases Sharpe ratios. (In fact, for short-horizon models there are fewer (4 or so) style factors.)} We further illustrate this via out-of-sample backtests using intraday mean-reversion alphas. In Section \ref{sec.6} we introduce risk models we refer to as ``heterotic CAPM", which are constructed similarly to the heterotic risk models, but the ``weights" for industry risk factors are based on style factors instead of principal components. In this novel, nontraditional implementation some style factors do add value. We briefly conclude in Section \ref{sec.7}. Appendix A contains R source code,\footnote{\, The source code given in Appendix \ref{app.A} is not written to be ``fancy" or optimized for speed or in any other way. Its sole purpose is to illustrate the algorithms described in the main text in a simple-to-understand fashion. Important legalese relating to this code is given in Appendix \ref{app.B}.} which implements our algorithm for general risk models, the complete Russian-doll embedding, style factor computation, etc.

\section{Multifactor Risk Models}\label{sec.2}
\subsection{Sample Covariance Matrix}

{}So, we have $N$ stocks and the associated time series of $N$ returns $R_i(t_s)$, $i=1,\dots,N$, where $t_s$, $s=0,1,2,\dots,M$ ($s=0$ corresponds to the most recent time), are the $M+1$ points in time in our time series (i.e., we have $M+1$ observations for each return). E.g., we could consider daily close-to-close returns, albeit the frequency is not critical here. Let $C_{ij}$ be the $N\times N$ sample covariance matrix computed based on these time series of returns. When $M<N$, the sample covariance matrix $C_{ij}$ is singular with $M$ positive eigenvalues.\footnote{\, The positive eigenvalue count can be lower if some returns are 100\% (anti-)correlated.} In this case we cannot invert $C_{ij}$, which is required in optimization (mean-variance optimization \cite{Q74}, Sharpe ratio maximization \cite{Sharpe94}, etc.). Furthermore, unless $M\gg N$, which is almost never (if ever) the case in practical applications, the off-diagonal elements of $C_{ij}$ (covariances) generally are not expected to be stable out-of-sample. In contrast, the diagonal elements (variances) typically are much more stable out-of-sample and can be relatively reliably computed even for $M\ll N$ (which, in fact, is often the case in practical applications). So, we need to replace the sample covariance matrix $C_{ij}$ by another {\em constructed} matrix -- call it $\Gamma_{ij}$ -- that is much more stable out-of-sample and invertible (positive-definite). That is, we must build a risk model.

\subsection{Total Variances}

{}In this regard, a lot of headache (see below) can be avoided by making the following simple observation. The main goal of a risk model is to predict the covariance matrix out-of-sample as precisely as possible, including the out-of-sample variances. However, even though this requirement is often overlooked in practical applications, a well-built risk model had better reproduce the in-sample variances. That is, we require that the risk model variances $\Gamma_{ii}$ coincide with the in-sample variances $C_{ii}$:
\begin{equation}\label{tot.risk}
 \Gamma_{ii} = C_{ii}
\end{equation}
Furthermore, as mentioned above, the $N$ variances $C_{ii}$ are relatively stable out-of-sample. It is therefore (a half of) the $N(N-1)$ off-diagonal covariances, which are generally unstable out-of-sample, we must actually model. Put differently, we must model the {\em correlations} $\Psi_{ij}$, $i\neq j$, where $\Psi_{ij}= C_{ij}/\sqrt{C_{ii}}\sqrt{C_{jj}}$ is the sample correlation matrix, whose diagonal elements  $\Psi_{ii} \equiv 1$. So, we need to replace the sample correlation matrix by another constructed matrix -- call it ${\widetilde \Gamma}_{ij}$ -- that is much more stable out-of-sample and invertible (positive-definite) subject to the conditions
\begin{equation}\label{diag.psi}
 {\widetilde \Gamma}_{ii} = \Psi_{ii} \equiv 1
\end{equation}
Once we build ${\widetilde\Gamma}_{ij}$, the risk model covariance matrix is given by $\Gamma_{ij} = \sqrt{C_{ii}}\sqrt{C_{jj}}~{\widetilde \Gamma}_{ij}$.

\subsection{Factor Models}

{}A popular way of constructing $\Gamma_{ij}$ is via a factor model. However, here we wish to construct ${\widetilde\Gamma}_{ij}$, not $\Gamma_{ij}$. Nonetheless, the factor model construction applies just as well. Instead of the returns $R_i(t_s)$ we simply use the ``normalized'' returns ${\widetilde R}_i(t_s)= R_i(t_s)/\sqrt{C_{ii}}$. Then $\Psi_{ij}$ is the sample covariance matrix of these normalized returns: $\Psi_{ij} = \mbox{Cov}({\widetilde R}_i(t_s), {\widetilde R}_j(t_s)) = \mbox{Cor}(R_i(t_s), R_j(t_s))$, where $\mbox{Cov}(\cdot, \cdot)$ and $\mbox{Cor}(\cdot, \cdot)$ are serial. I.e., we build a factor model for $\Psi_{ij}$ by treating ${\widetilde R}_i(t_s)$ as returns.

{}Then, we have $N$ random processes $\Upsilon_i$, which model the returns ${\widetilde R}_i$ (we omit $t_s$ dependence for notational convenience):
\begin{eqnarray}\label{Upsilon}
 &&\Upsilon_i = \chi_i + \sum_{A=1}^K \Omega_{iA}~f_A\\
 &&\mbox{Cov}(\chi_i, \chi_j) = \Xi_{ij} = \xi_i^2 ~\delta_{ij}\\
 &&\mbox{Cov}(\chi_i, f_A) = 0\\
 &&\mbox{Cov}(f_A, f_B) = \Phi_{AB}\\
 &&\mbox{Cov}(\Upsilon_i, \Upsilon_j) = {\widetilde \Gamma}_{ij}
\end{eqnarray}
where (in matrix notation)
\begin{equation}
 {\widetilde\Gamma} = \Xi + \Omega~\Phi~\Omega^T
\end{equation}
and $\delta_{ij}$ is the Kronecker delta; $\xi_i$ is the specific risk (a.k.a. idiosyncratic risk) for each stock; $\Omega_{iA}$ is an $N\times K$ factor loadings matrix; and $\Phi_{AB}$ is a $K\times K$ factor covariance matrix, $A,B=1,\dots,K$, where $K\ll N$. I.e., the random processes $\Upsilon_i$ are modeled via $N$ random processes $\chi_i$ (specific risk) together with $K$ random processes $f_A$ (factor risk). Assuming all $\xi_i>0$ and $\Phi_{AB}$ is positive-definite, then ${\widetilde \Gamma}_{ij}$ is automatically positive-definite (and invertible). Furthermore, ${\widetilde \Gamma}_{ij}$ is expected to be much more stable out-of-sample than $\Psi_{ij}$ as the number of risk factors, for which the factor covariance matrix $\Phi_{AB}$ needs to be computed, is $K\ll N$.

{}Since ${\widetilde R}_i$ are normalized, so are the specific risk and factor risk; indeed, (\ref{diag.psi}) gives the following $N$ conditions\footnote{\, With additional assumptions not all of these conditions are independent (see below).} for $N+K(K+1)/2$ unknowns:
\begin{equation}\label{diag.xi}
 \xi_i^2 + \sum_{A,B=1}^K \Omega_{iA}~\Phi_{AB}~\Omega_{iB} \equiv 1
\end{equation}
There are no ``natural'' $K(K+1)/2$ conditions we can impose in terms of out-of-sample unstable off-diagonal correlations $\Psi_{ij}$, $i\neq j$. So, we need additional assumptions to compute $\xi_i$ and $\Phi_{AB}$. {\em Intuitively}, one such assumption is that the total risk should be attributed to the factor risk to the greatest extent possible, i.e., the part of the total risk attributed to the specific risk should be minimized.

\subsection{Linear Regression}

{}One way to formulate this requirement mathematically is via least squares. Thus, mimicking (\ref{Upsilon}), we decompose the returns ${\widetilde R}_i$ via a linear model
\begin{equation}
 {\widetilde R}_i = \epsilon_i + \sum_{A=1}^K \Omega_{iA}~f_A
\end{equation}
Here the residuals $\epsilon_i$ are {\em not} the same as $\chi_i$ in (\ref{Upsilon}); in particular, generally the covariance matrix $\mbox{Cov}(\epsilon_i, \epsilon_j)$ is not diagonal (see below). We can require that
\begin{equation}\label{lin.unit}
 \sum_{i=1}^N \epsilon_i^2 \rightarrow \mbox{min}
\end{equation}
where the minimization is w.r.t. $f_A$. This produces a linear regression.\footnote{\, Without the intercept, that is, unless the intercept is already subsumed in $\Omega_{iA}$. Also, this is a regression with unit weights. Since the returns ${\widetilde R}_i$ are normalized, there are no ``natural'' out-of-sample stable candidates for the regression weights excepting the unit weights. This is the beauty of factoring out the sample volatilities $\sqrt{C_{ii}}$: the sample correlation matrix $\Psi_{ij}$ knows nothing about volatility, hence -- naturally -- unit weights. Also, note that for a large enough universe the sample volatilities have a skewed -- roughly log-normal -- distribution cross-sectionally. In contrast, $\Psi_{ii}$ are nicely uniform. This makes (\ref{diag.xi}) easier to satisfy (see below).} The solution to (\ref{lin.unit}) is given by (in matrix notation)
\begin{eqnarray}\label{fac.ret}
 &&f = \left({\Omega}^T~{\Omega}\right)^{-1}{\Omega}^T~{\widetilde R}\\
 &&{\epsilon} = \left[1 - Q\right] {\widetilde R}\\
 &&Q = {\Omega}\left({\Omega}^T~{\Omega}\right)^{-1}{\Omega}^T
\end{eqnarray}
where $Q = Q^T$ is a projection operator: $Q^2 = Q$. Consequently, we have:
\begin{eqnarray}
 &&{\widehat \Xi} = \mbox{Cov}\left({\epsilon},{\epsilon}^T\right) = \left[1 - Q\right] \Psi \left[1 - Q\right]\\
 &&{\Omega}~\Phi~{\Omega}^T = {\Omega}~\mbox{Cov}\left(f, f^T\right){\Omega}^T = Q~\Psi~Q
\end{eqnarray}
Note that the matrix ${\widehat \Xi}$ is not diagonal. However, the idea is to identify ${\xi}_i^2$ with the diagonal part of ${\widehat \Xi}$:
\begin{equation}\label{xi}
 {\xi}_i^2 = {\widehat \Xi}_{ii} = \left(\left[1 - Q\right] \Psi \left[1 - Q\right]\right)_{ii}
\end{equation}
and we have
\begin{equation}\label{gamma.twiddle}
 {\widetilde \Gamma}_{ij} = {\xi}_i^2~\delta_{ij} + \left(Q~\Psi~Q\right)_{ij}
\end{equation}
Note that ${\xi}_i^2$ defined via (\ref{xi}) are automatically positive (nonnegative, to be precise -- see below). However, we must satisfy the conditions (\ref{diag.psi}), which reduce to
\begin{eqnarray}\label{T}
 &&X_{ii} = 0\\
 &&X = 2~Q~\Psi~Q - Q~\Psi - \Psi~Q
\end{eqnarray}
The $N$ conditions (\ref{T}) are not all independent. Thus, we have $\mbox{Tr}(X) = 0$.

\subsection{Principal Components}\label{sub.pc}

{}The conditions (\ref{T}) are nontrivial. They are not satisfied for an arbitrary factor loadings matrix ${\Omega}_{iA}$. Thus, consider the simplest case of a single risk factor ($K=1$), so that the loadings matrix $\Omega_{iA}$ consists of a single column, call it $\omega_i$. Without loss of generality we can normalize it such that $\sum_{i=1}^N \omega_i^2=1$. Further, without loss of generality we can assume that all $\omega_i\neq 0$ -- indeed, for any vanishing $\omega_i$ factor risk vanishes and we only have specific risk, the matrix ${\widetilde \Gamma}_{ij}$ is block-diagonal, and we can ignore the block corresponding to the vanishing $\omega_i$. Then (\ref{T}) reduces to
\begin{equation}
 \sum_{j=1}^N \Psi_{ij}~\omega_j = \omega_i\sum_{j,k=1}^N \Psi_{jk}~\omega_j~\omega_k
\end{equation}
This invariably implies that $\omega_i$ is an eigenvector (principal component) of $\Psi_{ij}$.

\subsubsection{Multiple Principal Components}

{}More generally, (\ref{T}) is automatically satisfied if we take the columns of the factor loadings matrix to be, say, the first $K$ of the $N$ principal components $V_i^{(a)}$, $a=1,\dots,N$, of $\Psi_{ij}$ forming an orthonormal basis
\begin{eqnarray}\label{Psi.eigen}
 &&\sum_{j=1}^N \Psi_{ij}~V_j^{(a)} = \lambda^{(a)}~V_i^{(a)}\\
 &&\sum_{i=1}^N V_i^{(a)}~V_i^{(b)} = \delta_{ab}
\end{eqnarray}
such that the eigenvalues $\lambda^{(a)}$ are ordered decreasingly: $\lambda^{(1)} > \lambda^{(2)} >\dots$. More precisely, some eigenvalues may be degenerate. For simplicity -- and this is not critical here -- we will assume that all positive eigenvalues are non-degenerate. However, we can have multiple null eigenvalues. Typically, the number of nonvanishing eigenvalues\footnote{\, This number can be smaller if some stock returns are 100\% correlated or anti-correlated. For the sake of simplicity -- and this not critical here -- we will assume that there are no such returns.} is $M$, where, as above, $M+1$ is the number of observations in the stock return time series. We can readily construct a factor model with $K < M$:\footnote{\, For $K = M$ we have ${\widetilde \Gamma} = \Psi$, which is singular. Null eigenvalues do not contribute into (\ref{PC}).}
\begin{equation}\label{FLM.PC}
 {\Omega}_{iA} = \sqrt{\lambda^{(A)}}~V_i^{(A)},~~~A=1,\dots,K
\end{equation}
Then the factor covariance matrix $\Phi_{AB} = \delta_{AB}$ and we have
\begin{eqnarray}\label{PC}
 && {\widetilde \Gamma}_{ij} = {\xi}_i^2~\delta_{ij} + \sum_{A=1}^K \lambda^{(A)}~V_i^{(A)}~V_j^{(A)}\\
 && {\xi}_i^2 = 1 - \sum_{A=1}^K \lambda^{(A)}\left(V_i^{(A)}\right)^2
\end{eqnarray}
so ${\widetilde \Gamma}_{ii} = \Psi_{ii} \equiv 1$. See \cite{Het} for source code for building a risk model based on principal components (including for fixing the value of $K$).\footnote{\, Note that ${\xi}_i^2 = \sum_{a=K+1}^M \lambda^{(a)}\left(V_i^{(a)}\right)^2 \geq 0$. We are assuming $\lambda^{(a)}\geq 0$, which (up to computational precision) is the case if there are no N/As in the stock return times series.}

{}An evident limitation of the principal component approach is that the number of risk factors is limited by $M$. If long lookbacks are unavailabe/undesirable, as, e.g., in short-holding quantitative trading strategies, then typically $M \ll N$. Yet, the number of the actually relevant underlying risk factors can be substantially greater than $M$, and most of these risk factors are missed by this approach.\footnote{\, Another limitation is that the principal components are based on off-diagonal elements of $\Psi_{ij}$ and tend to be unstable out-of-sample, the first principal component typically being the most stable. In many applications one uses only the first principal component (see below).} Thus, it is unclear what to do with the principal components with null eigenvalues. They do not contribute to any sample factor covariance matrix. We need additional input.

\section{Industry Risk Factors}\label{sec.3}

{}While the number of risk factors based on principal components is limited, especially for shorter lookbacks,\footnote{\, The number of style factors is also limited, of order 10 or fewer (see below).} risk factors based on a granular enough industry classification can be ubiquitous. Furthermore, they are independent of the pricing data -- industry classification is based on fundamental/economic data (such as companies' products and services and more generally their revenue sources, suppliers, competitors, partners, etc.) -- and, in this regard, are essentially insensitive to the lookback. In fact, typically industry classification based risk factors tend to be rather stable out-of-sample as companies seldom jump industries (let alone sectors).

{}For terminological definiteness, here we will use the BICS\footnote{\, Bloomberg Industry Classification System.} nomenclature for the levels in the industry classification, albeit this is not critical here. Also, BICS has three levels ``sector $\rightarrow$ industry $\rightarrow$ sub-industry" (where ``sub-industry" is the most granular level). The number of levels in the industry hierarchy is not critical here either. So, we have: $N$ stocks labeled by $i=1,\dots,N$; $K$ sub-industries labeled by $A=1,\dots,K$; $F$ industries labeled by $a=1,\dots,F$;\footnote{\, The subscript index $a$ should not be confused with the superscript $(a)$ in (\ref{Psi.eigen}).} and $L$ sectors labeled by $\alpha=1,\dots,L$. More generally, we can think of such groupings as ``clusters". Sometimes, loosely, we will refer to such ``cluster" based factors as ``industry" factors.\footnote{\, Albeit in the BICS context we may be referring to, e.g., sub-industries, while in other classification schemes the actual naming may be altogether different.}

\subsection{``Binary" Property}

{}The binary property implies that each stock belongs to one and only one sub-industry, industry and sector (or, more generally, ``cluster"). Let $G$ be the map between stocks and sub-industries, $S$ be the map between sub-industries and industries, and $W$ be the map between industries and sectors:
\begin{eqnarray}\label{G.map}
 &&G:\{1,\dots,N\}\mapsto\{1,\dots,K\}\\
 &&S:\{1,\dots,K\}\mapsto\{1,\dots,F\}\label{S.map}\\
 &&W:\{1,\dots,F\}\mapsto\{1,\dots,L\}\label{W.map}
\end{eqnarray}
The beauty of the binary property is that the ``clusters" (sub-industries, industries and sectors) can be used to identify blocks (sub-matrices) in the correlation matrix $\Psi_{ij}$. E.g., for sub-industries the binary matrix $\delta_{G(i), A}$ defines such blocks.

\subsection{Binary Models}

{}Consider the following factor loadings matrix:
\begin{equation}\label{ind.bin}
 \Omega_{iA} = {1\over \sqrt{N_A}}~\delta_{G(i), A}
\end{equation}
where $J(A)= \{i|G(i)=A\}$ is the set of tickers (whose number $N(A)= |J(A)|$) in the sub-industry labeled by $A$. However, the factor loadings (\ref{ind.bin}) do not satisfy the conditions (\ref{T}). I.e., we cannot identify the factor covariance matrix with the sample covariance matrix of the regression coefficients and the specific risks with the variances of the regression residuals. There is a way around this difficulty. What if we continue to identify the factor covariance matrix with the sample covariance matrix of the regression coefficients and {\em define} the specific risks via (as opposed to (\ref{xi}))
\begin{equation}\label{xi.def}
 \xi_i^2 = 1 - \left( Q~\Psi~Q\right)_{ii}
\end{equation}
so we still have (\ref{gamma.twiddle}) and (\ref{diag.psi})? The issue with the definition (\ref{xi.def}) is that for a general factor loadings matrix $\Omega_{iA}$ some of the resulting $\xi_i^2$ unacceptably are {\em negative}. However, for the binary factor loadings (\ref{ind.bin}) $\xi_i^2$ are nonnegative. Indeed, we have
\begin{equation}
 \xi_i^2 = 1 - {1\over N^2(G(i))}\sum_{k,l\in J(G(i))} \Psi_{kl} \geq 1 - {1\over N^2(G(i))}\sum_{k,l\in J(G(i))} |\Psi_{kl}| \geq 0
\end{equation}
In fact, since we are assuming that we have no 100\% (anti-)correlated pairs of stocks, the only way we can have $\xi_i^2 = 0$ is for single-ticker sub-industries ($N(A) = 1$). These, in fact, are not problematic (see below). So, for binary factor loadings we can construct a factor model this way; however, with a caveat: if $M < K$, which is often the case, the factor covariance matrix
\begin{equation}
  \Phi_{AB} = {1\over\sqrt{N_A N_B}} \sum_{i\in J(A)}\sum_{j\in J(B)} \Psi_{ij}
\end{equation}
is singular. We will address this issue below via the Russian-doll risk model construction. But first let us discuss heterotic risk models \cite{Het}.

\subsection{Heterotic Models}\label{sub.het}

{}Consider the following factor loadings matrix:
\begin{eqnarray}\label{ind.pc}
 &&{\Omega}_{iA} = \delta_{G(i), A}~U_i\\
 &&U_i = [U(A)]_i,~~~i\in J(A),~~~A=1,\dots,K
\end{eqnarray}
where the $N(A)$-vector $U(A)$ is the {\em first} principal component of the $N(A) \times N(A)$ matrix $\Psi(A)$ defined via $[\Psi(A)]_{ij} = \Psi_{ij}$, $i,j\in J(A)$. (Note that $\sum_{i\in J(A)} [U(A)]_i^2 =1$; also, let the corresponding (largest) eigenvalue of $\Psi(A)$ be $\lambda(A)$.)\footnote{\, If $N(A)=1$, i.e., we have only one ticker in the sub-industry labeled by $A$, then $[U(A)]_i =1$ and $\lambda(A) = \Psi_{ii} = 1$, $i\in J(A)$.} With this factor loadings matrix we can compute the factor covariance matrix and specific risk via the above linear regression:\footnote{\, For single-ticker sub-industries ($N(A) = 1$) the specific risk vanishes: ${\xi}^2_i=0$; however, this does not pose a problem as this does not cause the matrix ${\widetilde\Gamma}_{ij}$ to be singular (see below).}
\begin{eqnarray}
 &&{\xi}_i^2 = 1 - \lambda(G(i))~ U_i^2\\
 &&{\widetilde \Gamma}_{ij} = \left[1 - \lambda(G(i))~ U_i^2\right]\delta_{ij} + U_i ~U_j \sum_{k\in J(G(i))}~ \sum_{l\in J(G(j))} U_k ~\Psi_{kl}~ U_l
\end{eqnarray}
and (\ref{T}) is automatically satisfied. This simplicity is due to the use of the (first) principal components corresponding to the blocks $\Psi(A)$ of the sample correlation matrix.\footnote{\, In fact, we can generalize this construction by using multiple principle components per ``cluster"; however, out-of-sample instability of the higher principal components mentioned above poses an issue. See \cite{Het} for details.} However, here too, as in the binary models, for short lookbacks we have $M < K$, so (note that $\Phi_{AA} = \lambda(A)$) the factor covariance matrix
\begin{equation}
 \Phi_{AB} = \sum_{i\in J(A)}~\sum_{j\in J(B)} U_i~\Psi_{ij}~U_j
\end{equation}
is singular. This again brings us to the nested Russian-doll risk model construction we discuss below. However, first we need to introduce one more ingredient.

\section{How to Deal with General Risk Factors?}\label{sec.4}

{}In heterotic risk models, and also when the factor loadings matrix is just a collection of principal components, the conditions (\ref{T}) are automatically satisfied. In binary risk models (\ref{T}) is not satisfied, but we found a workaround by simply defining the specific risk via (\ref{xi.def}). However, in more general cases this trick does not do. E.g., if we combine heterotic or binary risk factors with some style factors (see below) and/or principal components, (\ref{T}) is not satisfied, and (\ref{xi.def}) generally will produce $\xi_i^2$ at least some of which are negative. However, not all is lost. Thanks to modeling the {\em correlation} matrix $\Psi_{ij}$ via ${\widetilde \Gamma}_{ij}$ (as opposed to the covariance matrix $C_{ij}$ via $\Gamma_{ij}$), if we stick to the definition (\ref{xi}) via the linear regression, then (\ref{T}) is not satisfied for general $\Omega_{iA}$, but the violations of the condition (\ref{diag.psi}) are of order 1 as opposed to some large numbers.\footnote{\, Had we tried to do the same in the context of $\Gamma_{ij}$, the violations of the condition (\ref{tot.risk}) would be much more dramatic and nonuniform owing to the fact that the variances $C_{ii}$ have a highly skewed (quasi-log-normal) distribution. It is therefore more convenient to model $\Psi_{ij}$ via ${\widetilde \Gamma}_{ij}$.} Then we can simply rescale the matrix ${\widetilde \Gamma}_{ij}$ such that (\ref{diag.psi}) is satisfied, i.e., we can define ${\widetilde\Gamma}_{ij}$ via (as opposed to (\ref{gamma.twiddle}))
\begin{eqnarray}
 &&{\widetilde \Gamma}_{ij} = {1\over\gamma_i\gamma_j}\left[{\xi}_i^2~\delta_{ij} + \sum_{A,B=1}^K \Omega_{iA}~\Phi_{AB}~\Omega_{jB}\right] =  {1\over\gamma_i\gamma_j}\left[{\xi}_i^2~\delta_{ij} + \left(Q~\Psi~Q\right)_{ij}\right]\\
 &&\gamma_i^2 = {\xi}_i^2 + \left(Q~\Psi~Q\right)_{ii}
\end{eqnarray}
where, as before, $\xi_i^2$ is defined via (\ref{xi}) and is expressly nonnegative. This is a simple but powerful prescription that does the job: now we have (\ref{diag.psi}). (The {\em actual} factor loadings and specific risks are ${\widehat \Omega}_{iA}=\Omega_{iA}/\gamma_i$ and ${\widehat \xi}_i=\xi_i/\gamma_i$.) However, we still need to deal with the factor covariance matrix $\Phi_{AB}$ being singular for short lookbacks.

\section{Russian-Doll Construction}\label{sec.5}
\subsection{General Idea}

{}The simple idea behind the Russian-doll construction is to model such $\Phi_{AB}$ itself via yet another factor model matrix $\Gamma^\prime_{AB}$ (as opposed to computing it as a sample covariance matrix of the risk factor returns $f_A$):\footnote{\, We use a prime on ${\widetilde \Gamma}^\prime_{AB}$, $\gamma^\prime_A$, ${\xi}^{\prime}_A$, $\Phi_{ab}^\prime$, etc., to avoid confusion with ${\widetilde \Gamma}_{ij}$, $\gamma_i$, ${\xi}_i$, $\Phi_{AB}$, etc.}
\begin{eqnarray}
 &&\Gamma^\prime_{AB} = \sqrt{\Phi_{AA}}\sqrt{\Phi_{BB}}~{\widetilde \Gamma}^\prime_{AB}\\
 &&{\widetilde \Gamma}^\prime_{AB} = {1\over\gamma^\prime_A\gamma^\prime_B}\left[({\xi}^{\prime}_A)^2~\delta_{AB} + \sum_{a,b=1}^F {\Omega}^\prime_{Aa}~\Phi^\prime_{ab}~{\Omega}^\prime_{Bb}\right]\\
 &&(\gamma^\prime_A)^2 = ({\xi}^{\prime}_A)^2 + \sum_{a,b=1}^F {\Omega}^\prime_{Aa}~\Phi^\prime_{ab}~{\Omega}^\prime_{Ab}
\end{eqnarray}
where ${\xi}^\prime_A$ is the specific risk for the ``normalized" factor return ${\widetilde f}_A = f_A/\sqrt{\Phi_{AA}}$; ${\Omega}^\prime_{Aa}$, $A=1,\dots,K$, $a=1,\dots,F$ is the corresponding factor loadings matrix; and $\Phi^\prime_{ab}$ is the factor covariance matrix for the underlying risk factors $f^\prime_a$, $a=1,\dots,F$, where we assume that $F\ll K$. If the smaller factor covariance matrix $\Phi^\prime_{ab}$ is still singular, we model it via yet another factor model with fewer risk factors, and so on -- until the resulting factor covariance matrix is nonsingular. If, at the final stage, we are left with a single factor, then the resulting $1\times 1$ factor covariance matrix is automatically nonsingular -- it is simply the sample variance of the remaining factor.

\subsubsection{When We Have Industry Factors Only}

{}Moving forward we will need to distinguish between industry and non-industry\footnote{\, Here ``non-industry'' means ``not based on the industry classification''.}  factors. Examples of the latter are style factors and principal components. When we have only industry factors, we can use the natural hierarchy in the industry classification to implement the Russian-doll construction. For concreteness we will use the BICS terminology for the levels in the industry classification, albeit this is not critical here. Also, BICS has three levels ``$\mbox{sector}\rightarrow \mbox{industry} \rightarrow \mbox{sub-industry}$'' (where ``sub-industry'' is the most granular level). For definiteness, we will assume three levels here, albeit the generalization to more levels is straightforward. So, in the BICS terminology, we can use sub-industries as the risk factors for stocks, industries as the risk factors for sub-industries, sectors as the risk factors for industries, and (if need be) the ``market'' as the single risk factor for sectors (Kakushadze, 2015c,d). That is, a binary industry classification provides a natural nested hierarchy of risk factors. The binary models (\ref{ind.bin}) and the heterotic models (\ref{ind.pc}) discussed above both contain only industry factors.\footnote{\, In the heterotic risk models the columns of the factor loadings matrix are ``weighted'' by the first principal components in the sub-industries, industries, sectors and (if need be) ``market''.}

\subsubsection{Including Non-Industry Factors}

{}When we have both industry and non-industry factors, things are trickier. Thus, for non-industry risk factors it is not always as straightforward to identify a nested hierarchy of risk factors. E.g., for the principal component based risk factors there is no evident guiding principle to do so. However, not all is lost. In practice, the number of relevant style factors typically is substantially smaller than the number of industry factors (especially for (ultra-)short horizon models \cite{4F}), of order 10 or fewer. The number of principal components one may wish to include is also limited, either because of a short lookback, or out-of-sample instability of the higher principal components, or both. Thanks to this, we can start with a few non-industry (style and/or principal component) factors plus ubiquitous industry factors (typically, $\sim 100$ or more), and build a Russian-doll risk model.

{}The idea here is simple: there is no need to reduce the number of already-few non-industry factors, only that of the ubiquitous industry factors. That is, we apply the sequential nested Russian-doll embedding to the industry factors only, leaving the non-industry factors intact. At the final stage, we will have to compute the factor covariance matrix for the remaining factors, which will include all non-industry factors (along with, e.g., the sectors or the ``market'').

{}We will use the mid-Greek symbols $\mu, \nu,\dots$ to label the non-industry risk factors, and we use the $i, A, a, \alpha$ labels as above to label stocks, sub-industries, industries and sectors. Let ${\widetilde A} = (A, \mu)$, ${\widetilde a} = (a, \mu)$ and ${\widetilde \alpha} = (\alpha, \mu)$. Let $Y$ be the number of non-binary risk factors. Let ${\widetilde K}= K + Y$, ${\widetilde F}= F+Y$ and ${\widetilde L}= L+Y$. We will also use the index ${\widetilde \mu} =(0,\mu)$ with ${\widetilde Y} = Y+1$ values, where ``0'' labels the ``market''.

{}We then have:
\begin{eqnarray}
 && \Omega_{i{A}} = \omega_i~\delta_{G(i), A}\label{FLM0}\\
 && \Omega_{i{\widetilde A}} = \left(\Omega_{i{A}}, \Omega_{i\mu}\right)\\
 && \Omega^\prime_{Aa} = \omega^\prime_A~\delta_{S(A), a}\label{FLM1}\\
 && \Omega^\prime_{{\widetilde A}{\widetilde a}} = \mbox{diag}\left(\Omega^\prime_{Aa}, \delta_{\mu\nu}\right)\\
 && \Omega^{\prime\prime}_{a\alpha} = \omega^{\prime\prime}_a~\delta_{W(a), \alpha}\label{FLM2}\\
 && \Omega^{\prime\prime}_{{\widetilde a}{\widetilde \alpha}} = \mbox{diag}\left(\Omega^{\prime\prime}_{a\alpha}, \delta_{\mu\nu}\right)\\
 && \Omega^{\prime\prime\prime}_{\alpha 0} = \omega^{\prime\prime\prime}_\alpha\label{FLM3}\\
 && \Omega^{\prime\prime\prime}_{{\widetilde \alpha}{\widetilde \mu}} = \mbox{diag}\left(\Omega^{\prime\prime\prime}_{\alpha 0}, \delta_{\mu\nu}\right)
\end{eqnarray}
where: the maps $G$ (tickers to sub-industries), $S$ (sub-industries to industries) and $W$ (industries to sectors) are defined in (\ref{G.map}), (\ref{S.map}) and (\ref{W.map}); the columns $\Omega_{i\mu}$ of the factor loadings matrix $\Omega_{i{\widetilde A}}$ correspond to the non-industry factors; and we will specify the ``weights" $\omega_i$, $\omega^\prime_A$, $\omega^{\prime\prime}_a$ and $\omega^{\prime\prime\prime}_\alpha$ for the industry factors below.

\subsection{Complete Russian-Doll Embedding}\label{sub.het.rd}

{}We will continue to use the BICS terminology for the levels in the industry classification, albeit this is not critical here. Also, BICS has three levels ``sector $\rightarrow$ industry $\rightarrow$ sub-industry" (where ``sub-industry" is the most granular level). For definiteness, we will assume three levels here, and the generalization to more levels is straightforward. A nested Russian-doll risk model then is constructed as follows:
\begin{eqnarray}\label{Gamma.RD}
 &&\Gamma_{ij} = \sqrt{C_{ii}}\sqrt{C_{jj}}~{\widetilde \Gamma}_{ij}\\
 &&{\widetilde \Gamma}_{ij} = {\widehat \xi}_i^2~\delta_{ij} + \sum_{{\widetilde A},{\widetilde B}=1}^{\widetilde K} {\widehat \Omega}_{i{\widetilde A}}~\Gamma^\prime_{{\widetilde A}{\widetilde B}}~{\widehat \Omega}_{j{\widetilde B}}\\
 &&\Gamma^\prime_{{\widetilde A}{\widetilde B}} = \sqrt{\Phi_{{\widetilde A}{\widetilde A}}}\sqrt{\Phi_{{\widetilde B}{\widetilde B}}}~{\widetilde \Gamma}^\prime_{{\widetilde A}{\widetilde B}}\label{Gamma.prime.RD}\\
 &&{\widetilde \Gamma}^\prime_{{\widetilde A}{\widetilde B}} = ({\widehat \xi}^{\prime}_{\widetilde A})^2~\delta_{{\widetilde A}{\widetilde B}} + \sum_{{\widetilde a},{\widetilde b}=1}^{\widetilde F} {\widehat \Omega}^\prime_{{\widetilde A}{\widetilde a}}~\Gamma^{\prime\prime}_{{\widetilde a}{\widetilde b}}~{\widehat \Omega}^\prime_{{\widetilde B}{\widetilde b}}\\
&&\Gamma^{\prime\prime}_{{\widetilde a}{\widetilde b}} = \sqrt{\Phi^\prime_{{\widetilde a}{\widetilde a}}}\sqrt{\Phi^\prime_{{\widetilde b}{\widetilde b}}}~{\widetilde \Gamma}^{\prime\prime}_{{\widetilde a}{\widetilde b}}\\
 &&{\widetilde \Gamma}^{\prime\prime}_{{\widetilde a}{\widetilde b}} = ({\widehat \xi}^{\prime\prime}_{\widetilde a})^2~\delta_{{\widetilde a}{\widetilde b}} + \sum_{{\widetilde \alpha},{\widetilde \beta}=1}^{\widetilde L} {\widehat \Omega}^{\prime\prime}_{{\widetilde a}{\widetilde \alpha}}~\Gamma^{\prime\prime\prime}_{{\widetilde \alpha}{\widetilde \beta}}~{\widehat \Omega}^{\prime\prime}_{{\widetilde b}{\widetilde \beta}}\\
 &&\Gamma^{\prime\prime\prime}_{{\widetilde \alpha}{\widetilde \beta}} = \sqrt{\Phi^{\prime\prime}_{{\widetilde \alpha}{\widetilde \alpha}}}\sqrt{\Phi^{\prime\prime}_{{\widetilde \beta}{\widetilde \beta}}}~{\widetilde \Gamma}^{\prime\prime\prime}_{{\widetilde \alpha}{\widetilde \beta}}\\
 &&{\widetilde \Gamma}^{\prime\prime\prime}_{{\widetilde \alpha}{\widetilde \beta}} = ({\widehat \xi}^{\prime\prime\prime}_{\widetilde \alpha})^2~\delta_{{\widetilde
 \alpha}{\widetilde \beta}} + \sum_{{\widetilde \mu},{\widetilde \nu}=1}^{\widetilde Y} {\widehat \Omega}^{\prime\prime\prime}_{{\widetilde \alpha}{\widetilde \mu}}~\Phi^{\prime\prime\prime}_{{\widetilde \mu}{\widetilde \nu}}~{\widehat \Omega}^{\prime\prime\prime}_{{\widetilde \mu}{\widetilde \nu}}
\end{eqnarray}
where
\begin{eqnarray}\label{xi.RD}
 &&{\widehat \xi}_i = \xi_i /\gamma_i\\
 &&{\widehat \Omega}_{i{\widetilde A}} = {\Omega}_{i{\widetilde A}} / \gamma_i\label{Omega.RD}\\
 &&\gamma_i^2 = {\xi}_i^2 + \left(Q~\Psi~Q\right)_{ii}\\
 &&\xi_i^2 = \left(\left[1 - Q\right]\Psi\left[1 - Q\right]\right)_{ii}\\
 &&Q = \Omega \left(\Omega^T \Omega\right)^{-1}\Omega^T\label{Q0}\\
 &&{\widehat \xi}^\prime_{\widetilde A} = \xi^\prime_{\widetilde A} /\gamma^\prime_{\widetilde A}\\
 &&{\widehat \Omega}^\prime_{{\widetilde A}{\widetilde a}} = {\Omega}^\prime_{{\widetilde A}{\widetilde a}} / \gamma^\prime_{\widetilde A}\\
 &&(\gamma^\prime_{\widetilde A})^2 = ({\xi}^{\prime}_{\widetilde A})^2 + \left(Q^\prime~\Psi^\prime~Q^\prime\right)_{{\widetilde A}{\widetilde A}}\\
 &&({\xi}^\prime_{\widetilde A})^2 = \left(\left[1 - Q^\prime\right]\Psi^\prime\left[1 - Q^\prime\right]\right)_{{\widetilde A}{\widetilde A}}\\
 &&Q^\prime = \Omega^\prime \left(\Omega^{\prime T}\Omega^\prime\right)^{-1}\Omega^{\prime T}\label{Q1}\\
 &&{\widehat \xi}^{\prime\prime}_{\widetilde a} = \xi^{\prime\prime}_{\widetilde a} /\gamma^{\prime\prime}_{\widetilde a}\\
 &&{\widehat \Omega}^{\prime\prime}_{{\widetilde a}{\widetilde \alpha}} = {\Omega}^{\prime\prime}_{{\widetilde a}{\widetilde \alpha}} / \gamma^{\prime\prime}_{\widetilde a}\\
 &&(\gamma^{\prime\prime}_{\widetilde a})^2 = ({\xi}^{\prime\prime}_{\widetilde a})^2 + \left(Q^{\prime\prime}~\Psi^{\prime\prime}~Q^{\prime\prime}\right)_{{\widetilde a}{\widetilde a}}\\
 &&({\xi}^{\prime\prime}_{\widetilde a})^2 = \left(\left[1 - Q^{\prime\prime}\right]\Psi^{\prime\prime}\left[1 - Q^{\prime\prime}\right]\right)_{{\widetilde a}{\widetilde a}}\\
 &&Q^{\prime\prime} = \Omega^{\prime\prime} \left(\Omega^{\prime\prime T}\Omega^{\prime\prime}\right)^{-1}\Omega^{\prime\prime T}\label{Q2}\\
 &&{\widehat \xi}^{\prime\prime\prime}_{\widetilde \alpha} = \xi^{\prime\prime\prime}_{\widetilde \alpha} /\gamma^{\prime\prime\prime}_{\widetilde \alpha}\\
 &&{\widehat \Omega}^{\prime\prime\prime}_{{\widetilde \alpha}{\widetilde \mu}} = {\Omega}^{\prime\prime\prime}_{{\widetilde \alpha}{\widetilde \mu}} / \gamma^{\prime\prime\prime}_{\widetilde \alpha}\\
 &&(\gamma^{\prime\prime\prime}_{\widetilde \alpha})^2 = ({\xi}^{\prime\prime\prime}_{\widetilde \alpha})^2 + \left(Q^{\prime\prime\prime}~\Psi^{\prime\prime\prime}~Q^{\prime\prime\prime}\right)_{{\widetilde \alpha}{\widetilde \alpha}}\\
 &&({\xi}^{\prime\prime\prime}_{\widetilde \alpha})^2 = \left(\left[1 - Q^{\prime\prime\prime}\right]\Psi^{\prime\prime\prime}\left[1 - Q^{\prime\prime\prime}\right]\right)_{{\widetilde \alpha}{\widetilde \alpha}}\\
 &&Q^{\prime\prime\prime} = \Omega^{\prime\prime\prime} \left(\Omega^{\prime\prime\prime T}\Omega^{\prime\prime\prime}\right)^{-1}\Omega^{\prime\prime\prime T}\label{Q3}
\end{eqnarray}
In (\ref{Q0}) $\Omega$ is the $N\times {\widetilde K}$ matrix ${\Omega}_{i{\widetilde A}}$; in (\ref{Q1}) $\Omega^\prime$ is the ${\widetilde K} \times {\widetilde F}$ matrix $\Omega^\prime_{{\widetilde A}{\widetilde a}}$; in (\ref{Q2}) $\Omega^{\prime\prime}$ is the ${\widetilde F} \times {\widetilde L}$ matrix $\Omega^{\prime\prime}_{{\widetilde a}{\widetilde \alpha}}$; and in (\ref{Q3}) $\Omega^{\prime\prime\prime}$ is the ${\widetilde L} \times {\widetilde Y}$ matrix $\Omega^{\prime\prime\prime}_{{\widetilde \alpha}{\widetilde \mu}}$. Also,
\begin{eqnarray}
 &&\Phi = \left(\Omega^T~\Omega\right)^{-1} \Omega^T \Psi~\Omega \left(\Omega^T~\Omega\right)^{-1}\\
 &&\Phi^\prime = \left(\Omega^{\prime T}\Omega^\prime\right)^{-1} \Omega^{\prime T} \Psi^\prime~\Omega^\prime \left(\Omega^{\prime T}\Omega^\prime\right)^{-1}\\
 &&\Phi^{\prime\prime} = \left(\Omega^{\prime\prime T}\Omega^{\prime\prime}\right)^{-1} \Omega^{\prime\prime T} \Psi^{\prime\prime}~\Omega^{\prime\prime} \left(\Omega^{\prime\prime T}\Omega^{\prime\prime}\right)^{-1}\label{Theta.1}\\
 &&\Phi^{\prime\prime\prime} = \left(\Omega^{\prime\prime\prime T}\Omega^{\prime\prime\prime}\right)^{-1} \Omega^{\prime\prime\prime T} \Psi^{\prime\prime\prime}~\Omega^{\prime\prime\prime} \left(\Omega^{\prime\prime\prime T}\Omega^{\prime\prime\prime}\right)^{-1}\\
 &&\Psi_{ij} = C_{ij} / \sqrt{C_{ii}}\sqrt{C_{jj}}\\
 &&\Psi^\prime_{{\widetilde A}{\widetilde B}} = \Phi_{{\widetilde A}{\widetilde B}} / \sqrt{\Phi_{{\widetilde A}{\widetilde A}}}\sqrt{\Phi_{{\widetilde B}{\widetilde B}}}\\
 &&\Psi^{\prime\prime}_{{\widetilde a}{\widetilde b}} = \Phi^\prime_{{\widetilde a}{\widetilde b}} / \sqrt{\Phi^\prime_{{\widetilde a}{\widetilde a}}}\sqrt{\Phi^\prime_{{\widetilde b}{\widetilde b}}}\\
 &&\Psi^{\prime\prime\prime}_{{\widetilde \alpha}{\widetilde \beta}} = \Phi^{\prime\prime}_{{\widetilde \alpha}{\widetilde \beta}} / \sqrt{\Phi^{\prime\prime}_{{\widetilde \alpha}{\widetilde \alpha}}}\sqrt{\Phi^{\prime\prime}_{{\widetilde \beta}{\widetilde \beta}}}
\end{eqnarray}
Finally, the ``weights" $\omega_i$, $\omega^\prime_A$, $\omega^{\prime\prime}_a$ and $\omega^{\prime\prime\prime}_\alpha$ in (\ref{FLM0}), (\ref{FLM1}), (\ref{FLM2}) and (\ref{FLM3}) are give by
\begin{eqnarray}\label{U.RD}
 &&\omega_i = [U(A)]_i,~~~i\in J(A)\\
 &&\omega^\prime_A = [U^\prime(a)]_A,~~~A\in J^\prime(a)\\
 &&\omega^{\prime\prime}_a = [U^{\prime\prime}(\alpha)]_a,~~~a\in J^{\prime\prime}(\alpha)
\end{eqnarray}
Here $J(A) = \{i| G(i)=A\}$ ($N_A = |J(A)|$ tickers in sub-industry $A$), $J^\prime(a) = \{A|S(A) = a\}$ ($N^\prime(a) = |J^\prime(a)|$ sub-industries in industry $a$), $J^{\prime\prime}(\alpha) = \{a|W(a) = \alpha\}$ ($N^{\prime\prime}(\alpha)= |J^{\prime\prime}(\alpha)|$ industries in sector $\alpha$). The ``market'' contains all $N$ tickers. The $N(A)$-vector $U(A)$ is the first principal component of $\Psi(A)$ with the eigenvalue $\lambda(A)$ ($[\Psi(A)]_{ij} = \Psi_{ij}$, $i,j\in J(A)$); the $N^\prime(a)$-vector $U^\prime(a)$ is the first principal component of $\Psi^\prime(a)$ with the eigenvalue $\lambda^\prime(a)$ ($[\Psi^\prime(a)]_{AB} = \Psi^\prime_{AB}$, $A,B\in J^\prime(a)$); the $N^{\prime\prime}(\alpha)$-vector $U^{\prime\prime}(\alpha)$ is the first principal component of $\Psi^{\prime\prime}(\alpha)$ with the eigenvalue $\lambda^{\prime\prime}(\alpha)$ ($[\Psi^{\prime\prime}(\alpha)]_{ab} = \Psi^{\prime\prime}_{ab}$, $a,b\in J^{\prime\prime}(\alpha)$); finally, $\omega^{\prime\prime\prime}_\alpha$ is the first principal component of $\Psi^{\prime\prime\prime}_{\alpha\beta}$ with the eigenvalue $\lambda^{\prime\prime\prime}$. The vectors $U(A)$, $U^\prime(a)$ and $U^{\prime\prime}(\alpha)$ are normalized, so $\sum_{i\in J(A)} U_i^2 = 1$, $\sum_{A\in J^\prime(a)} (U^\prime_A)^2 = 1$, $\sum_{a\in J^{\prime\prime}(\alpha)} (U^{\prime\prime}_a)^2 = 1$, and also $\sum_{\alpha=1}^L (\omega^{\prime\prime\prime}_\alpha)^2 = 1$.

{}For the sake of completeness, above we included the step where the sample factor covariance matrix $\Phi^{\prime\prime}_{{\widetilde \alpha}{\widetilde \beta}}$ for the sectors plus non-industry factors is further approximated via a ${\widetilde Y}$-factor model ${\widetilde \Gamma}^{\prime\prime\prime}_{{\widetilde \alpha}{\widetilde \beta}}$, where ${\widetilde Y} = Y + 1$, i.e., the $L$ risk factors corresponding to the sectors are modeled via a 1-factor model (the  ``market"), while the $Y$ non-industry risk factors are untouched (see above). If $\Phi^{\prime\prime}_{{\widetilde \alpha}{\widetilde \beta}}$ computed via (\ref{Theta.1}) is nonsingular, then this last step can be omitted,\footnote{\, That is, assuming there are enough observations in the time series for out-of-sample stability.} so at the last stage we have ${\widetilde L} = L + Y$ factors (as opposed to ${\widetilde Y}$ factors).\footnote{\, For the sake of completeness, the definitions of the factor returns at each stage are as follows (in matrix notation): (i) for the sub-industry plus non-industry factor returns $f_{\widetilde A}$: $f = \left(\Omega^T~\Omega\right)^{-1} \Omega^T {\widetilde R}$, where ${\widetilde R}_i = R_i/\sqrt{C_{ii}}$; (ii) for the industry plus non-industry factor returns $f^\prime_{\widetilde a}$: $f^\prime = \left(\Omega^{\prime T}\Omega^\prime\right)^{-1} \Omega^{\prime T} {\widetilde f}$, where ${\widetilde f}_{\widetilde A} = f_{\widetilde A}/\sqrt{\Phi_{{\widetilde A}{\widetilde A}}}$; (iii) for the sector plus non-industry factor returns $f^{\prime\prime}_{\widetilde \alpha}$: $f^{\prime\prime} = \left(\Omega^{\prime\prime T}\Omega^{\prime\prime}\right)^{-1} {\widetilde f}^\prime$, where ${\widetilde f}^\prime_{\widetilde a} = f^\prime_{\widetilde a}/\sqrt{\Phi^\prime_{{\widetilde a}{\widetilde a}}}$; and (iv) for the ``market" plus non-industry factor returns $f^{\prime\prime\prime}_{\widetilde\mu}$: $f^{\prime\prime\prime} = \left(\Omega^{\prime\prime\prime T}\Omega^{\prime\prime\prime}\right)^{-1} \Omega^{\prime\prime\prime T} {\widetilde f}^{\prime\prime}$, where ${\widetilde f}^{\prime\prime}_{\widetilde \alpha} = f^{\prime\prime}_{\widetilde \alpha}/\sqrt{\Phi^{\prime\prime}_{{\widetilde \alpha}{\widetilde \alpha}}}$.} Similarly, if we have enough observations to compute the sample covariance matrix $\Phi^\prime_{{\widetilde a}{\widetilde b}}$ for the industries plus non-industry factors, we can stop at that stage. Finally, note that in the above construction we are guaranteed to have $({\widetilde \xi}^{\prime\prime\prime}_\alpha)^2 > 0$, $({\widetilde \xi}^{\prime\prime}_a)^2 >0$, $({\widetilde \xi}^\prime_A)^2> 0$ and ${\widetilde \xi}_i^2\geq 0$. The last equality occurs only for single-ticker sub-industries (see below).

\subsection{Model Covariance Matrix and Its Inverse}\label{sub.inv}

{}The model covariance matrix is given by $\Gamma_{ij}$ defined in (\ref{Gamma.RD}). For completeness, let us present it in the ``canonical" form:
\begin{equation}
 \Gamma_{ij} = {\widetilde \xi}_i^2~\delta_{ij} + \sum_{{\widetilde A},{\widetilde B}=1}^{\widetilde K} {\widetilde\Omega}_{i{\widetilde A}}~\Phi^*_{{\widetilde A}{\widetilde B}}~{\widetilde \Omega}_{j{\widetilde B}}
\end{equation}
where
\begin{eqnarray}\label{actual.xi}
 &&{\widetilde \xi}_i^2 = C_{ii}~{\widehat \xi}_i^2\\
 &&{\widetilde \Omega}_{i{\widetilde A}} = \sqrt{C_{ii}}~{\widehat \Omega}_{i{\widetilde A}}\label{actual.Omega}\\
 &&\Phi^*_{{\widetilde A}{\widetilde B}} = \Gamma^\prime_{{\widetilde A}{\widetilde B}}
\end{eqnarray}
where ${\widehat \xi}_i^2$ is defined in (\ref{xi.RD}), $U_i$ is defined in (\ref{Omega.RD}), $\Gamma^\prime_{{\widetilde A}{\widetilde B}}$ is defined in (\ref{Gamma.prime.RD}), and we use the star superscript in the factor covariance matrix $\Phi^*_{{\widetilde A}{\widetilde B}}$ (which is nonsingular) to distinguish it from the sample factor covariance matrix $\Phi_{{\widetilde A}{\widetilde B}}$ (which is singular).

{}In many applications, such as portfolio optimization, one needs the inverse of the matrix $\Gamma$. When we have no single-ticker sub-industries, the inverse is given by (in matrix notation)
\begin{eqnarray}\label{Gamma.inv}
 &&\Gamma^{-1} = {\widetilde \Xi}^{-1} - {\widetilde \Xi}^{-1}~{\widetilde \Omega}~\Delta^{-1}~{\widetilde \Omega}^T~{\widetilde \Xi}^{-1}\\
 &&\Delta = (\Phi^*)^{-1} + {\widetilde \Omega}^T~{\widetilde \Xi}^{-1}~{\widetilde \Omega}\\
 &&{\widetilde \Xi} = \mbox{diag}({\widetilde \xi}^2_i)
\end{eqnarray}
However, when there are some single-ticker sub-industries, the corresponding ${\widetilde \xi}_i^2=0$, $i\in H$, where $H = \{i|N(G(i)) = 1\}$, so (\ref{Gamma.inv}) ``breaks". Here is a simple ``fix". We can rewrite $\Gamma_{ij}$ via
\begin{equation}
 \Gamma_{ij} = {\overline \xi}_i^2~\delta_{ij} + \sum_{{\widetilde A},{\widetilde B}=1}^{\widetilde K} {\widetilde \Omega}_{i{\widetilde A}}~{\overline \Phi}^*_{{\widetilde A}{\widetilde B}}~{\widetilde \Omega}_{j{\widetilde  B}}
\end{equation}
where: ${\overline \xi}_i^2 = {\widetilde \xi}_i^2$ for $i\not\in H$; ${\overline \xi}_i^2 = \sum_{A=1}^K {\widetilde \Omega}^2_{iA}~\Phi^*_{AA}$ for $i\in H$; and ${\overline \Phi}^*_{AA} = 0$ for $A\in E$, where $E = \{A|N(A)=1\}$, whereas for all other values of ${\widetilde A}$ and ${\widetilde B}$ we have ${\overline\Phi}^*_{{\widetilde A}{\widetilde B}} = \Phi^*_{{\widetilde A}{\widetilde B}}$. Now we can invert $\Gamma$ via
\begin{eqnarray}\label{Gamma.inv.1}
 &&\Gamma^{-1} = {\overline \Xi}^{-1} - {\overline \Xi}^{-1}~{\widetilde \Omega}~{\overline \Delta}^{-1}~{\widetilde \Omega}^T~{\overline \Xi}^{-1}\\
 &&{\overline\Delta} = ({\overline \Phi}^*)^{-1} + {\widetilde \Omega}^T~{\overline \Xi}^{-1}~{\widetilde \Omega}\\
 &&{\overline \Xi} = \mbox{diag}({\overline \xi}^2_i)
\end{eqnarray}
Note that, due to the factor model structure, to invert the $N\times N$ matrix $\Gamma$, we only need to invert two ${\widetilde K}\times {\widetilde K}$ matrices ${\overline \Phi}^*$ and ${\overline \Delta}$. If there are no single-ticker sub-industries, then $\Phi^*$ itself has a factor model structure and involves inverting two ${\widetilde F}\times {\widetilde F}$ matrices, one of which has a factor model structure, and so on.

\subsection{Do Non-industry Factors Add Value?}

{}Above we described an algorithm for constructing general multifactor risk models containing both industry and non-industry based risk factors. More precisely, for the industry based risk factors we chose the ``weights" $\omega_i$, $\omega^\prime_A$, $\omega^{\prime\prime}_a$ and $\omega^{\prime\prime\prime}_\alpha$ in (\ref{FLM0}), (\ref{FLM1}), (\ref{FLM2}) and (\ref{FLM3}) as the first principal components of the correlation (sub-)matrices for the corresponding ``clusters" (sub-industries, industries, sectors, ``market"). However, these ``weights" can be chosen at will, e.g., we could have chosen some or all of them to be binary (i.e., as in (\ref{ind.bin})). In fact, these choices need not even be uniform across different ``clusters", e.g., we could choose binary ``weights" for some sub-industries and first (or higher) principal components for other sub-industries. It is a separate issue whether the resulting model would perform better.

{}The question we address in this subsection is whether non-industry factors actually add value. An off-the-cuff answer might appear to be that they do, at least for style factors.\footnote{\, If the non-industry factors are based on principal components, one may have a priori doubts whether they add value considering out-of-sample stability issues (see above).} However, this ``intuition" based on longer-horizon models does us no good for short-lookback models we are after here. And there is a simple way to test this, to wit, by utilizing the same intraday mean-reversion alphas as in \cite{Het}. In the remainder of this subsection, which very closely follows most parts of Section 6 of \cite{Het},\footnote{\, We ``rehash" it here not to be repetitive but so that the presentation herein is self-contained.} we describe the backtesting procedure.

\subsubsection{Notations}

{}Let $P_{is}$ be the time series of stock prices, where $i=1,\dots,N$ labels the stocks, and $s=0,1,\dots,M$ labels the trading dates, with $s=0$ corresponding to the most recent date in the time series. The superscripts $O$ and $C$ (unadjusted open and close prices) and $AO$ and $AC$ (open and close prices fully adjusted for splits and dividends) will distinguish the corresponding prices, so, e.g., $P^C_{is}$ is the unadjusted close price. $V_{is}$ is the unadjusted daily volume (in shares). Also, for each date $s$ we define the overnight return as the previous-close-to-open return:
\begin{equation}\label{c2o.ret}
 E_{is} = \ln\left({P^{AO}_{is} / P^{AC}_{i,s+1}}\right)
\end{equation}
This return will be used in the definition of the expected return in our mean-reversion alpha. We will also need the close-to-close return
\begin{equation}\label{c2c.ret}
 R_{is} = \ln\left({P^{AC}_{is} / P^{AC}_{i,s+1}}\right)
\end{equation}
An out-of-sample (see below) time series of these returns will be used in constructing the risk models. All prices in the definitions of $E_{is}$ and $R_{is}$ are fully adjusted.

{}We assume that: i) the portfolio is established at the open\footnote{\, This is a so-called ``delay-0" alpha: the same price, $P^O_{is}$ (or adjusted $P^{AO}_{is}$), is used in computing the expected return (via $E_{is}$) and as the establishing fill price.} with fills at the open prices $P^O_{is}$; ii) it is liquidated at the close on the same day -- so this is a purely intraday alpha -- with fills at the close prices $P^C_{is}$; and iii) there are no transaction costs or slippage -- our aim here is not to build a realistic trading strategy, but to test {\rm relative} performance of various risk models and see what adds value to the alpha and what does not. The P\&L for each stock
\begin{equation}
 \Pi_{is} = H_{is}\left[{P^C_{is}\over P^O_{is}}-1\right]
\end{equation}
where $H_{is}$ are the {\em dollar} holdings. The shares bought plus sold (establishing plus liquidating trades) for each stock on each day are computed via $Q_{is} = 2 |H_{is}| / P^O_{is}$.

\subsubsection{Universe Selection}\label{sub.univ}

{}For the sake of simplicity,\footnote{\, In practical applications, the trading universe of liquid stocks typically is selected based on market cap, liquidity (ADDV), price and other (proprietary) criteria.} we select our universe based on the average daily dollar volume (ADDV) defined via (note that $A_{is}$ is out-of-sample for each date $s$):
\begin{equation}\label{ADDV}
 A_{is}= {1\over d} \sum_{r=1}^d V_{i, s+r}~P^C_{i, s+r}
\end{equation}
We take $d=21$ (i.e., one month), and then take our universe to be the top 2000 tickers by ADDV. To ensure that we do not inadvertently introduce a universe selection bias, we rebalance monthly (every 21 trading days, to be precise). I.e., we break our 5-year backtest period (see below) into 21-day intervals, we compute the universe using ADDV (which, in turn, is computed based on the 21-day period immediately preceding such interval), and use this universe during the entire such interval. We do have the survivorship bias as we take the data for the universe of tickers as of 9/6/2014 that have historical pricing data on http://finance.yahoo.com (accessed on 9/6/2014) for the period 8/1/2008 through 9/5/2014. We restrict this universe to include only U.S. listed common stocks and class shares (no OTCs, preferred shares, etc.) with BICS sector, industry and sub-industry assignments as of 9/6/2014.\footnote{\, The choice of the backtesting window is intentionally taken to be exactly the same as in \cite{Het} to simplify various comparisons, which include the results therefrom.} However, as discussed in detail in Section 7 of \cite{MeanRev}, the survivorship bias is not a leading effect in such backtests.\footnote{\, Here we are after the {\em relative outperformance}, and it is reasonable to assume that, to the leading order, individual performances are affected by the survivorship bias approximately equally as the construction of all alphas and risk models is ``statistical" and oblivious to the universe.}

\subsubsection{Backtesting}\label{sub.back}

{}We run our simulations over a period of 5 years (more precisely, 1260 trading days going back from 9/5/2014, inclusive). The annualized return-on-capital (ROC) is computed as the average daily P\&L divided by the intraday investment level $I$ (with no leverage) and multiplied by 252. The annualized Sharpe Ratio (SR) is computed as the daily Sharpe ratio multiplied by $\sqrt{252}$. Cents-per-share (CPS) is computed as the total P\&L (in cents, not dollars) divided by the total shares traded.\footnote{\, As mentioned above, we assume no transaction costs, which are expected to reduce the ROC of the optimized alphas by the same amount as all strategies trade the exact same amount by design. Therefore, including the transaction costs would have no effect on the actual {\em relative outperformance} in the horse race, which is what we are after here.}

\subsubsection{Optimized Alphas}\label{sub.opt}

{}The optimized alphas are based on the expected returns $E_{is}$ optimized via Sharpe ratio maximization using the risk models we are testing, i.e., the covariance matrix $\Gamma_{ij}$ given by (\ref{Gamma.RD}), which we compute every 21 trading days (same as for the universe). For each date (we omit the index $s$) we maximize the Sharpe ratio subject to the dollar neutrality constraint:
\begin{eqnarray}
 &&{\cal S} = {\sum_{i=1}^N H_i~E_i\over{\sqrt{\sum_{i,j=1}^N \Gamma_{ij}~H_i~H_j}}} \rightarrow \mbox{max}\\
 &&\sum_{i=1}^N H_i = 0\label{d.n.opt}
\end{eqnarray}
The solution is given by
\begin{equation}\label{H.opt}
 H_i = -\gamma \left[\sum_{j = 1}^N \Gamma^{-1}_{ij}~E_j - \sum_{j=1}^N \Gamma^{-1}_{ij}~{{\sum_{k,l=1}^N \Gamma^{-1}_{kl}~E_l}\over{\sum_{k,l = 1}^N \Gamma^{-1}_{kl}}}\right]
\end{equation}
where $\Gamma^{-1}$ is the inverse of $\Gamma$ (see Subsection \ref{sub.inv}), and $\gamma > 0$ (mean-reversion alpha) is fixed via (we set the investment level $I$ to \$20M in our backtests)
\begin{equation}
 \sum_{i=1}^N \left|H_i\right| = I
\end{equation}
Note that (\ref{H.opt}) satisfies the dollar neutrality constraint (\ref{d.n.opt}).

{}The simulation results are given in Table \ref{table1} for (i) the heterotic risk model based on BICS sectors only (i.e., we have only one level), (ii) the heterotic risk model based on BICS industries (i.e., we have two levels, sectors and industries), (iii) the heterotic risk model based on BICS sub-industries (i.e., we have three levels, sectors, industries and sub-industries), (iv) the heterotic risk model based on BICS sub-industries plus the ``prc" style risk factor of \cite{4F},\footnote{\, The corresponding column in the factor loadings matrix $\Omega_{i{\widetilde A}}$ is defined as $\ln(P^{AC}_{i,s+1})$ for the first (i.e., earliest) date labeled by $s$ in the aforesaid 21-day backtesting period and kept the same for the entire such period. Other similar definitions of this factor (such as averaging over, say, the 21-day period immediately preceding the backtesting period) give very similar results.}
(v) the heterotic risk model based on BICS sub-industries plus the ``hlv" style risk factor of \cite{4F}, (vi) the heterotic risk model based on BICS sub-industries plus the ``vol" style risk factor of \cite{4F},\footnote{\, The intraday volatility based ``hlv" and average volume based ``vol" factors are recomputed based on 21-day periods, same as the price based ``prc" factor. For the formulaic definitions of the ``prc", ``mom", ``hlv" and ``vol" factors see \cite{4F} and the {\tt{\small qrm.style()}} function in Appendix \ref{app.A} hereof, where the R code is essentially formulaic (i.e., reads like formulas).} (vii) the heterotic risk model based on BICS sub-industries plus the ``mom" style risk factor of \cite{4F},\footnote{\, For using the prior day's open-to-close momentum based risk factor (see below) to make sense, it is necessary to recompute the risk model every day, which is what -- unlike all other rows based on 21-day recalculations -- the ``mom" row in Table \ref{table1} shows. The drop in SR and CPS is largely due to the noise from daily recalculations. Thus, the heterotic risk model based on BICS sub-industries (case (iii) above) recomputed daily has ROC 55.94\%, SR 14.80 and CPS 2.66. The ``mom" factor for the date labeled by $s$ is defined as $\ln(P^{C}_{i,s+1}/P^O_{i,s+1})$ \cite{4F}.\label{fn.mom}} (viii) the heterotic risk model based on BICS sub-industries plus the ``prc", ``hlv" and ``vol" style risk factors,\footnote{\, We deliberately omit the ``mom" factor -- see footnote \ref{fn.mom}.} (ix) the heterotic risk model based on BICS sub-industries plus the first principal component of the sample correlation matrix $\Psi_{ij}$, (x) the heterotic risk model based on BICS sub-industries plus the first 10 principal components of the sample correlation matrix,\footnote{\, The principal components are computed out-of-sample based on the 21-day windows, same as for the heterotic (industry classification based) risk factors.} and (xi) the heterotic risk model based on BICS sub-industries plus the first principal component of the sample correlation matrix and the ``prc" style risk factor. The results in Table \ref{table1} unequivocally suggest that adding granularity to the industry classification adds sizable value. However, any improvement from style factors and/or principal components is at best marginal.

{}How come? The reason is simple. By construction, when we add a new column to the factor loadings matrix, only its part orthogonal to the already present columns contributes to the risk model -- indeed, the risk model is invariant under the linear transformations of the form $\Omega_{i{\widetilde A}}\rightarrow \sum_{{\widetilde B} = 1}^{\widetilde K}\Omega_{i{\widetilde B}}~D_{{\widetilde B}{\widetilde A}}$, where $D_{{\widetilde B}{\widetilde A}}$ is a nonsingular matrix. BICS sub-industries cover a large subspace of the (relevant)\footnote{\, While there is no such thing as a ``perfect" industry classification, well-constructed industry classifications do provide value in covering the risk space as they are based on ``economic" considerations, to wit, the products, services, revenue sources, competitors, partners, suppliers, clients, etc., of the companies they classify.} risk space. It is difficult for a few style risk factors and/or principal components to provide a new relevant direction not covered by a few hundred sub-industries. And the number of relevant style risk factors is at most of order 10 for longer-horizon models and even fewer (around 4) for shorter-lookback risk models. Prosaically, it boils down to a numbers game, which the ubiquitous industry factors win hands down...\footnote{\, For long horizons, cf. \cite{ind.lead}.}

\section{Heterotic CAPM}\label{sec.6}

{}Does this mean that the style risk factors are useless? Not quite. It is just that the way to use them is not to trivially append them to the much more numerous industry factors, which is how it is done traditionally. In this section we discuss a different way of utilizing the style factors, which does appear to add sizable value.

{}The idea -- or the thought process -- is quite simple. We know the industry factors add value. In fact, we know that adding granularity (i.e., going from sectors to industries to sub-industries) adds value. I.e., breaking up the universe of tickers into ``clusters" (sectors, industries, sub-industries) makes a lot of sense. So, why try to apply the style risk factors to the entire universe when we can apply them at the level of each ``cluster" (e.g., sub-industry)? This is analogous to what we did in the heterotic risk models. Instead of using principal components (of the sample correlation matrix) for the entire universe, we used them for each ``cluster". So, the idea is this. What if we use the style factors to define the ``weights" $\omega_i$ in (\ref{FLM0}) at the (most granular) level of sub-industries?

\subsection{One Factor per Sub-industry}\label{sub.capm.q}

{}In this subsection we discuss a construction, which we term ``Heterotic CAPM" for the reasons we explain below, which utilizes a single style factor, e.g., ``prc", to define the ``weights" $\omega_i$ in (\ref{FLM0}). We will generalize this construction to include multiple style factors below. Also, we will keep the ``weights" $\omega^\prime_A$, $\omega^{\prime\prime}_a$ and $\omega^{\prime\prime\prime}_\alpha$ in (\ref{FLM1}), (\ref{FLM2}) and (\ref{FLM3}) based on principal components. We could also utilize style factors at the levels of industries, sectors and the ``market", but we will not do so here.

{}Table \ref{table2} gives the backtest results for $\omega_i$ defined as (i) the intercept,\footnote{\, I.e., we take $\omega_i=\nu_i \equiv 1$. The intercept $\nu_i$ is just the simplest style factor.} (ii) ``prc", (iii) ``hlv", (iv) ``vol", and (v) ``mom".\footnote{\, In cases (i)-(iv) we recalculate the risk model every 21 days, and daily in case (v) (see above).} These results appear to be all over the place. And they are for a good reason. Let us focus on the ``prc" based model. Suppressing the time series index, we have $\omega_i = \ln(P^{AC}_i)$. However, there is absolutely no reason why we could not define $\omega_i = \ln(P^{AC}_i/\mu)$, where $\mu$ is some constant. I.e., $\omega_i$ is only defined up to a shift proportional to the intercept $\nu_i \equiv 1$: $\omega_i = \ln(P^{AC}_i) - \ln(\mu)~\nu_i$, and a priori there is no reason to set $\mu = 1$. The backtest results depend on this choice.\footnote{\, Thus, when $\mu = 1$, we get the second row in Table \ref{table2}. However, when $\mu$ is large (or close to zero), we get the first row in Table \ref{table2}. The normalizations (i.e., their means) of ``prc", ``hlv" and ``vol" are all very different, hence the disparate results in Table \ref{table2}. For ``mom" we have an additional effect from the extra noise due to daily recomputations.} So, what should this $\mu$ be? There is no magic bullet here. We simply have a 1-parameter family, and this parameter can be optimized based on backtests. However, neither is it guaranteed to be out-of-sample stable (i.e., it will have to be recompute frequently), nor is it guaranteed to be independent of the alpha model which is used in the backtesting. {\em As\'i es la vida}. In fact, it is even worse: we could define
$\omega_i = \ln(P^{AC}_i) - \ln(\mu_A)~\nu_i$, $i\in J(A)$, i.e., we can have a different normalization factor $\mu_A$ for each sub-industry labeled by $A$. However, that would be too many parameters to deal with, so let us stick to $\mu$ uniform over all sub-industries.

{}In this regard, it is convenient to parameterize the ``weights" $\omega_i$ as follows:
\begin{equation}\label{capm.q}
 \omega_i = q~\omega^*_i + \left(1-2q\right){\overline\omega}^*~\nu_i
\end{equation}
where $\omega^*_i = \ln(P^{AC}_i)$ and ${\overline \omega}^* = {1\over N}\sum_{j=1}^N\omega_j^*$. So, for $q=0$ we have the intercept (the overall normalization factor is immaterial), for $q=1$ we have demeaned $\omega_i = \omega^*_i - {\overline \omega}^*$, and for $q=1/2$ we have $\omega_i = \omega_i^*$. The simulation results for the values of $q$ ranging from 0 (which is equivalent to the intercept) to 0.5 (the style factor ``as-is") to 1.0 (the demeaned style factor) are given for ``prc" in Table \ref{table3} and ``hlv" in Table \ref{table4}. For ``vol" the best simulated performance is for $|q| \gg 1$ (ROC 56.72\%, SR 15.23, CPS 2.67) with inferior performance in between. It is evident that the demeaned style factor has inferior performance,\footnote{\, This is because roughly half of the values of a demeaned style factor are positive, and the rest are negative, so roughly half of the correlations within each sub-industry come out to be negative, which makes little sense financially considering stocks in the same sub-industry on average should be positively correlated. The same applies to the ``mom" factor, which (without demeaning) has roughly equal numbers of positive and negative values. More precisely, this is the case on days when the market is not running up or sliding down. In the heterotic CAPM models we have ${\widetilde \Gamma}_{ij} = {\widehat \xi}_i^2~\delta_{ij} + \omega_i~\omega_j~\Gamma^\prime_{G(i),G(j)}$, so for the pair-wise ($i\neq j$) correlations within the same sub-industry ($G(i)=G(j)$) we have $\mbox{sign}({\widetilde \Gamma}_{ij}) = \mbox{sign}\left(\omega_i~\omega_j\right)$, hence the aforesaid roughly 50-50 split.} so does the intercept, and the optimal value of $q$ corresponds to a combination of the two, except that this combination is not necessarily the same as $\omega_i^*$, i.e., it does not necessarily correspond to $q = 0.5$.

{}The theoretical reason for this is clear. The factor risk model construction is not invariant under the shifts $\Omega_{iA} \rightarrow \Omega_{iA} + \chi_A$, which implies that the ``centering" of the style factors matters -- the overall multiplicative normalization invariance due to the invariance under the homogeneous linear transformations $\Omega_{iA} \rightarrow \sum_{A=1}^K \Omega_{iB}~D_{BA}$, where $D_{BA}$ is a nonsingular matrix, does not extend inhomogeneously.

{}In any event, it is evident that the ``prc" style factor potentially adds most value by sizably improving CPS. However, to make sure we are not dealing with a spurious improvement, we must check whether it persists once we add position/liquidity bounds. Therefore, in Table \ref{table5} we give the simulation results with bounds for the vanilla heterotic and the ``prc" based heterotic CAPM risk models. The strict bounds are
\begin{equation}\label{liq}
 |H_{is}| \leq 0.01~A_{is}
\end{equation}
where $A_{is}$ is ADDV defined in (\ref{ADDV}). For the sake of simplicity we take $q=0.5$ (i.e., the ``prc" factor ``as-is", $\omega_i = \omega_i^*$) for the heterotic CAPM model as this choice, albeit not optimal, does not raise out-of-sample stability questions. We use the R code in Appendix C of \cite{Het} for optimization with bounds. Table \ref{table5} shows that, happily, the improvement in CPS persists even with the bounds (\ref{liq}).

\subsection{Multiple Factors per Sub-industry}

{}In the above construction we used only one style factor per sub-industry. We can readily generalize this construction to utilize multiple style factors. Suppose we have $Y$ style factors labeled by $\mu=1,\dots,Y$  and $K$ sub-industries. Na\"ively, we can take an $N \times (KY)$ factor loadings matrix of the form $\Omega_{i{\widehat A}} = \omega_{i\mu}\delta_{G(i), A}$, where ${\widehat A} = (A,\mu)$ takes $KY$ values, and $\omega_{i\mu}$ are the factor loadings for the style factors. However, this choice will not work for small sub-industries. We can try to do overly complicated things, but in practice it does not pay off. So, let us do something simple. First, let $Y_A$ be the number of style factors for the sub-industry $A$. So, $Y_A \leq Y$. Thus, if $N(A) < Y$, then we can simply have a single style factor ($Y_A=1$) for such a sub-industry and identify it with the first style factor in $\omega_{i\mu}$, i.e., we can order the columns in $\omega_{i\mu}$ such that the first column is the default column (e.g., ``prc"). If $N(A)\geq Y$, we can set $Y_A = \mbox{min}(N(A)-1, Y)$ and take the first $Y_A$ of the $Y$ style factors. We must also make sure that the resulting style factors $\omega_{i\mu}$, $\mu = 1,\dots,Y_A$, for each sub-industry are linearly independent (see the source code in Appendix \ref{app.A}). The simulated performance (without bounds) for, e.g., ``prc + hlv" ($Y=2$, ``prc" is the default) has the following characteristics: ROC 53.15\%, SR 16.39\%, CPS 2.54. The increase in the Sharpe ratio is not surprising -- there are substantially more risk factors. However, it comes at a price, to wit, the reduction in ROC and CPS.

\subsection{A Simple Test for Adding New Factors}\label{sub.kappa}

{}As we discussed above, it is understandable that appending a few style factors to the much more numerous sub-industries makes a little difference. However, using multiple style factors per each sub-industry as in the heterotic CAPM models does not appear to add much value either. How come? There are two parts to this story.

{}First, the entire idea -- which is essentially how most (if not all) ``fundamental" commercial multifactor risk models are built -- of appending style factors to industries is not all that justified. Here is why. Consider a 1-factor model for the correlation matrix: ${\widetilde \Gamma}_{ij} = {\xi}_i^2\delta_{ij} + \beta_i\beta_j$, where $\beta_i$ is the single column of our factor loadings matrix (and we have absorbed the $1\times 1$ factor covariance matrix into the definition of $\beta_i$). It makes a lot of sense to take $\beta_i$ to be proportional to the first principal component $V^{(1)}_i$ of the sample correlation matrix $\Psi_{ij}$. This is because $\Psi_{ij} = \lambda^{(1)}V^{(1)}_iV^{(1)}_j + \dots$, where the ellipses stand for the higher principal component terms, which are subleading when $N$ is large.\footnote{\, The out-of-sample stability of $V^{(1)}_i$ is a separate issue. For large $N$, in the leading approximation, we have $V^{(1)}_i\approx 1/\sqrt{N}$ (the so-called ``market mode"), which is evidently stable.} It further makes a lot of sense to do this for each sub-industry, i.e., model the correlation matrix $[\Psi(A)]_{ij}$ for each sub-industry labeled by $A$ via its first principal component -- and this is nothing but the heterotic risk model of \cite{Het} (see Subsection \ref{sub.het}). However, a priori there is no reason why any given style factor $\beta_i$ should be a good candidate for why the off-diagonal elements of the correlation matrix should be well-approximated by a bilinear tensor $\beta_i\beta_j$. The ``lore" for justifying using style factors as $\beta_i$ goes as follows. Suppose we take historical returns and regressed them over some style factors. If the correlations are sufficiently high (e.g., if we have high Fama-MacBeth (1973) t-statistic), then using such style factors in a multifactor model is justified. However, this argument has an evident caveat. A factor model -- by construction -- assumes that the residuals of that regression should have low correlations with the factor returns and the pair-wise correlations between different residuals should also be low. Furthermore, this should persist out-of-sample. None of this is guaranteed by simply having sufficiently high correlations between returns and style factors.

{}Second, a simple test for checking whether a style (or any other) factor may add value is to compute the following coefficient:
\begin{equation}\label{kappa}
 \kappa = {\sum_{i,j=1}^N \beta_i~\Psi_{ij}~\beta_j \over{\lambda^{(1)}\sum_{i=1}^N \beta_i^2}}
\end{equation}
As above, $\lambda^{(1)}$ is the largest eigenvalue of the sample correlation matrix $\Psi_{ij}$. By definition, $\kappa \leq 1$ (and $\kappa = 1$ only when $\beta_i = V_i^{(1)}$, up to an overall normalization factor). If $\kappa$ is small, or is not small in-sample, but is unstable out-of-sample (i.e., takes small values from sample to sample), then such $\beta_i$ is not going to add much value. In (\ref{kappa}), whenever possible, $\beta_i$ should be computed based on a time period prior to the time period based on which $\Psi_{ij}$ is computed. For short lookbacks, which we are primarily interested in here, usually this is possible.

{}We can rewrite (\ref{kappa}) as follows:
\begin{equation}\label{kappa.1}
 \kappa = \sum_{a = 1}^M {\lambda^{(a)}\over{\lambda^{(1)}}} \left[\sum_{i=1}^N \beta^\prime_i~V^{(a)}_i\right]^2
\end{equation}
where $\beta^\prime_i = \beta_i /\sqrt{\sum_{j=1}^N\beta_j^2}$. If $N\gg 1$, we usually have $\lambda^{(1)} \gg \lambda^{(2)}$, so assuming $\left|\beta^\prime_i~V^{(1)}_i\right|$ is not small, we have
\begin{equation}\label{kappa.2}
 \kappa \approx \left[\sum_{i=1}^N \beta^\prime_i~V^{(1)}_i\right]^2
\end{equation}
and $\kappa$ approximately measures the inner product (i.e., the cross-sectional ``correlation") of $\beta_i^\prime$ with the first principal component.

{}In fact, we can compute $\kappa$ for each sub-industry:
\begin{equation}\label{kappa.sub}
 \kappa(A) = {\sum_{i,j\in J(A)} \beta_i~[\Psi(A)]_{ij}~\beta_j \over{\lambda(A)\sum_{i\in J(A)} \beta_i^2}}
\end{equation}
where $\lambda(A)$ is the largest eigenvalue of $[\Psi(A)]_{ij}$ and, as above, $J(A)$ is the set of the values of the index $i$ in the sub-industry labeled by $A$. For sub-industries with small enough numbers of tickers $N(A)$ (compared with $M$), the sample correlation matrix $[\Psi(A)]_{ij}$ is reliable enough, and therefore so is $\kappa(A)$. For large sub-industries $[\Psi(A)]_{ij}$ is not as reliable and, in fact, is singular (when $M < N(A)$). However, the first principal component $[V(A)]_i$ of $[\Psi(A)]_{ij}$ is relatively reliable, and if $N(A)$ is large enough, we can apply the approximation (\ref{kappa.2}) to such sub-industries
\begin{equation}\label{kappa.3}
 \kappa(A) \approx \left[\sum_{i\in J(A)} \beta^\prime_i~[V(A)]_i\right]^2
\end{equation}
assuming that $\left|\sum_{i\in J(A)} \beta^\prime_i~[V(A)]_i\right|$ is not small (see below). The bottom line is that $\kappa$ and $\kappa(A)$ measure how much value $\beta_i$ is expected to add in a factor model.\footnote{\, Another way of approaching this issue is, as in \cite{KLT}, to pull the off-diagonal elements of $\Psi_{ij}$ into a vector and regress it over the similarly-pulled three tensors $\nu_i\nu_j$, $\nu_i\beta_j + \beta_i \nu_j$ and $\beta_i\beta_j$, where, as above, $\nu_i\equiv 1$ is the intercept. However, this regression makes sense only if $\Psi_{ij}$ itself is reliably computed, i.e., if $M\gg N$, which is not the case here. We could attempt to do this for small sub-industries, but this would only give a partial picture, so here we opt to work with the coefficient $\kappa$ instead. (In \cite{KLT} this regression was applied to alphas and $M$ was large enough compared with $N$.)}

{}Now, if $\left|\sum_{i\in J(A)} \beta^\prime_i~[V(A)]_i\right| \ll 1$, then this factor cannot possibly give a sizable improvement. We can always express $\beta^\prime_i$, $i\in J(A)$, as a linear combination of the eigenvectors of $[\Psi(A)]_{ij}$. We have no control over the eigenvectors with null eigenvalues, which are present for $M < N(A)$. We simply do not have enough data (owing to the short lookback) to deal with those directions in the risk space. As to the higher principal components with positive eigenvalues, their contribution to the performance is at best subleading and unstable out-of-sample. This is aptly illustrated by taking the vanilla heterotic model construction (Subsection \ref{sub.het}) and replacing the first principal component for each sub-industry by the $k_{pc}$-th principal component, where $k_{pc} > 1$. More precisely, we need to deal with small sub-industries, so for a fixed $k_{pc}$ we replace the first principal component for the sub-industry labeled by $A$ by the $k^\prime_{pc}$-th principal component, where $k^\prime_{pc} = \mbox{min}(k_{pc}, N(A))$. The result of the simulation for the values of $k_{pc}$ from 1 to 5 is given in Table \ref{table6}. These results are unequivocal: higher principal components by themselves are suboptimal. Why? For the same reason as why demeaned style factors are suboptimal: for large $N$ we know that $V^{(1)}_i\approx1/\sqrt{N}$, so $V^{(a)}_i$, $a > 1$, have approximately vanishing sums. The same holds for large sub-industries, and also roughly for the rest of the sub-industries (because the tickers within each sub-industry should be relatively highly correlated).\footnote{\, This statement should be understood statistically. Even for a well-built industry classification, sometimes tickers are ``misclassified", and even if they are properly classified, the correlations do not always aline with sub-industries, but {\em on average} they do, or else the vanilla heterotic risk model would not work as well as it does.} Consequently, as we discussed above, roughly half of the pair-wise correlations within each sub-industry are negative, which makes little sense.\footnote{\, On the flip side, if we replace the first principal component by the intercept, the performance also worsens (see above). This is because in this case we have uniform correlations within each sub-industry. The first principal component is close to the rescaled intercept but still captures the variability in the pair-wise correlations, hence better performance.}

{}So, in plain English, the game of picking a style (or whatever other) factor is in trying to find a linear combination of the first principal component (with a sizable weight) and some higher principal components in such a way that it is as stable out-of-sample as possible. This is a tough game to play, especially when the number of observations (and, therefore, of the available principal components) is limited. So, style factors are simply shortcuts or rough {\em intelligent guesses} based on some financial/economic considerations. However, they do not necessarily work well in the factor model context. E.g., using long-horizon style factors such as value or growth for short-horizon risk models (for use with trading strategies with holding periods of order of a few days or intraday) makes little sense \cite{CustomRM}, and even factors such as size (log of the market cap) must be stripped off of long-horizon elements (for size one strips off the shares outstanding and arrives at log of the price) \cite{4F}. However, it goes beyond that. The results in Table \ref{table6} indicate that it is just a single linear combination that is expected to add value, not multiple style factors. This is consistent with our results above, that ``prc" adds most value, albeit this does not imply that ``prc" is the most optimal combination, in fact, above we saw that it is not. However, there is no simple or ``natural" way to determine an out-of-sample stable ``optimal" combination.

{}As they say, a picture is worth a thousand words. In Figure 1 we plot the value of the coefficient $\kappa(A)$ vs. $N(A)$, where $\kappa(A)$ is defined in (\ref{kappa.sub}). We emphasize that the computation of $\kappa(A)$ in Figure 1 is: (i) 100\% out-of-sample (see below), and (ii) wholly independent of any trading strategy, alpha, etc. To compute it, we do the following. We take the same backtesting period as in Subsections \ref{sub.univ} and \ref{sub.back}, break it up into 21 day periods as above, take 2,000 top tickers by ADDV for each period, calculate the correlation matrix $[\Psi(A)]_{ij}$ for each sub-industry\footnote{\, More precisely, we drop small sub-industries with fewer than 4 tickers.} for each such period, and calculate the ``prc", ``mom", ``hlv" and ``vol" style factors based on a 21-day period\footnote{\, The ``prc" factor is computed based on the day immediately preceding said 21-day period (averaging over the entire preceding 21-day period makes little difference). The ``mom" factor normally would be computed as an intraday open-to-close return based on the day immediately preceding said 21-day period; however, for this test we define ``mom" as a 21-day moving average of such intraday open-to-close returns to smooth out the noise (albeit this turns out to make a little difference). The ``hlv" and ``vol" factors are defined as above, via 21-day moving averages.} immediately preceding such period (so everything is out-of-sample). Figure 1 combines data for all such 21-day periods. As expected, ``mom" is most unstable, while ``prc", ``hlv" and ``vol" have similar behavior, $\kappa(A)$ is close to 1 for large sub-industries (as these style factors have a large intercept component), and for smaller industries it is less stable. Again, the bottom line is that the intercept (or the first principal component) is the {\em Konzertmeister} here.\footnote{\, For the sake of completeness, let us note that if $\beta_i$ corresponds to a sub-industry factor of the form (\ref{ind.bin}), assuming $V^{(1)}_i\approx 1/\sqrt{N}$, we have $\kappa \approx N(A)/N \ll 1$, where $N(A)$ is the number of tickers in the sub-industry and we are assuming that the number of sub-industries is large and $N(A) \ll N$. I.e., in this case it takes a combination of essentially all or most sub-industries to get a sizable contribution into the factor model -- individual sub-industries have subleading contributions. As they say, there is strength in numbers...}

\section{Concluding Remarks}\label{sec.7}

{}So -- at long last -- in this paper we give a complete algorithm and source code for constructing general multifactor risk models (for equities) via any combination of style factors, principal components and/or industry factors. For short lookbacks, so that the factor covariance matrix is nonsingular, we employ the Russian-doll risk model construction \cite{RusDoll}. This generalizes the heterotic risk model construction \cite{Het} to include arbitrary non-industry risk factors as well as industry risk factors with generic ``weights". When these ``weights" for the sub-industry risk factors are based on a style factor (as opposed to principal components), we refer to such a construction as heterotic CAPM. The name is due to the fact that for each sub-industry we have a 1-factor model analogous to CAPM as the factor loadings are based on a style factor. In the case of the original CAPM it is (log of) the market cap; in the case of short lookbacks we have log of the price instead (the ``prc" style factor \cite{4F}), albeit in principle it can be any other style factor such as ``hlv" or ``vol" or even log of the market cap.\footnote{\, Log of the market cap would make sense for long horizons. For short horizons ``prc" appears to add most value (compared with the ``hlv", ``vol" and ``mom" factors).}

{}As we discussed above, generally appending a few style (or principal component) factors to the much more ubiquitous industry factors (numbering in a few hundred at the most granular level) adds little value. This is because well-constructed industry classifications such as BICS or GICS\footnote{\, Global Industry Classification Standard.} already capture a lot of the relevant risk space at their most granular levels. So, a way to squeeze value from the style factors is to use them as the ``weights" in the industry factors as in our heterotic CAPM models. However, even there the improvement is not earthshaking: thus, the best performing ``prc" factor based heterotic CAPM does sizably improve CPS, but at the expense of lowering ROC and SR a bit. Overall, for short-horizon models the heterotic risk model construction appears to be a safe and stable choice.\footnote{\, As we discussed above, there is no simple, ``natural" way to pick the ``centering" of a style factor such as ``prc" or ``hlv", i.e., where its mean should be. Furthermore, any given choice is not guaranteed to be out-of-sample stable. In this regard, using the first principal components as the ``weights" for the industry risk factors as in the heterotic construction is ``natural", albeit even the first principal components are not 100\% out-of-sample stable. However, unlike higher principal components, they are sufficiently stable, hence the stability of the heterotic risk models.}

{}Another remark concerns the factor models generally. When the specific risk is computed via (\ref{xi}),\footnote{\, That in general we must rescale it as in Section \ref{sec.4} does not alter the point we make here.} there is some degree of the inherent out-of-sample instability in the sample correlation matrix $\Psi_{ij}$ that seeps into $\xi_i^2$ = $([1-Q]\Psi[1-Q])_{ii}$ via the $Q\Psi$ and $\Psi Q$ terms. In this regard, note that the bilinear in $Q$ term $Q\Psi Q = \Omega\Phi \Omega^T$ is different, because it is expressed via the (much more stable) factor covariance matrix $\Phi_{AB}$. The linear in $Q$ terms $Q\Psi$ and $\Psi Q$ cannot be rewritten via $\Phi_{AB}$, they are related to the correlations between stock returns and factor returns, which are not as stable as factor correlations. One way to mitigate this instability is by using the definition $\xi_i^2 = 1 - (Q\Psi Q)_{ii}$ instead. However, as we discussed above, this definition is not guaranteed to yield positive $\xi_i^2$. One way to deal with this is to ``squash" the factor risk as in the {\tt{\small qrm.fac()}} function in Appendix \ref{app.A}. However, this ``squashing" itself introduces its own instabilities with a worse net result (see Appendix \ref{app.A}).

{}So, now that we have provided a complete algorithm and source code for constructing general multifactor risk models, we hope quantitative traders will build their own risk models as opposed to wasting money on off-the-shelf alternatives, which are not optimized for (short-horizon) quantitative trading but are mostly geared toward standardized risk management in the context of the longer-horizon investments (mutual/pension funds, etc.) \cite{CustomRM}. The usual excuse that some hard-to-get data is required to build such models is just that, an excuse. Most serious quant traders already have all the data required to build a stable short-horizon risk model, such as price-volume and industry classification data. And complicated ``fundamental" data going back many years is not required as it is not relevant at short-horizons \cite{CustomRM}.\footnote{\, Nor is the implied volatility data, as its relevance at short horizons is expected to be at best marginal, and there is no empirical evidence that it adds value for short-horizon strategies \cite{Het}. (Cf. an older paper (Ederington and Guan, 2002), which sometimes is referred to in the context of using implied volatility for ``fundamental" (long-horizon) risk models.)} So, stop wasting money and complaining, start building risk models and enjoy!

\appendix

\section{R Code: General Russian-Doll Risk Model}\label{app.A}
{}In this appendix we give the R (R Package for Statistical Computing, http://www.r-project.org) source code for building general risk models, which has options for building the heterotic risk models, heterotic CAPM models (based on a single or multiple style factors), and models with the traditional treatment of non-industry (style and/or principal component) risk factors (whereby they are appended to the industry based risk factors). See Sections \ref{sec.2}-\ref{sec.6} for details. The code below is essentially self-explanatory and straightforward as it simply follows the formulas therein.

{}The function {\tt{\small qrm.cov.gen(ret, load, calc.inv = T)}} calculates the covariance matrix $\Gamma_{ij} = \sigma_i\sigma_j{\widetilde\Gamma}_{ij}$ ($\sigma_i = \sqrt{C_{ii}}$) for a general factor loadings matrix $\Omega_{iA}$ using the algorithm of Section \ref{sec.4}. The input is as follows: {\tt{\small ret}} is an $N\times d$ matrix of returns $R_{is}$ (e.g., daily close-to-close returns), where $N$ is the number of tickers, $d = M+1$ is the number of observations in the time series (e.g., the number of trading days), and the ordering of the dates is immaterial; ii) {\tt{\small load}} is the $N\times K$ factor loadings matrix $\Omega_{iA}$; if {\tt{\small calc.inv = T}} the code also computes and returns the inverse of $\Gamma_{ij}$; otherwise it does not. The output is a list: {\tt{\small result\$spec.risk}} is the actual specific risk ${\widetilde \xi_i} = \sigma_i\xi_i/\gamma_i$, {\tt{\small result\$fac.load}} is the actual factor loadings matrix ${\widetilde \Omega}_{iA} = \sigma_i\Omega_{iA}/\gamma_i$ (see Section \ref{sec.4}), {\tt{\small result\$fac.cov}} is the factor covariance matrix $\Phi_{AB}$, {\tt{\small result\$cov.mat}} is the factor model covariance matrix $\Gamma_{ij}$, {\tt{\small result\$inv.cov}} is the matrix $\Gamma^{-1}_{ij}$ inverse to $\Gamma_{ij}$ (populated only if {\tt{\small calc.inv = T}}), and {\tt{\small result\$fac.ret}} is the $K \times d$ matrix of factor returns $f_{As}$. If $M < N$, the factor loadings matrix is singular (so the inverse matrix $\Gamma^{-1}_{ij}$ cannot be computed). This is dealt with by employing the Russian-doll embedding, which is implemented via the next function.

{}The function {\tt{\small qrm.gen.het(ret, ind, mkt.fac = F, rm.sing.tkr = F, p = 0,\phantom{\,}
append.style = F, k.style = 0, style = 0)}} reduces to the function {\tt{\small qrm.het(ret, ind, mkt.fac = F, rm.sing.tkr = F)}} in Appendix B of \cite{Het} for the default values {\tt{\small p = 0, append.style = F, k.style = 0, style = 0}}. The input is: i) {\tt{\small ret}}, which is the same as above; ii) {\tt{\small ind}}, a list whose length a priori is arbitrary, and its elements are populated by the binary matrices (with rows corresponding to tickers, so {\tt{\small dim(ind[[$\cdot$]])[1]}} is $N$) corresponding to the levels in the input binary industry classification hierarchy in the order of decreasing granularity (for BICS {\tt{\small ind[[1]]}} is the $N\times K$ matrix $\delta_{G(i), A}$ (sub-industries), {\tt{\small ind[[2]]}} is the $N\times F$ matrix $\delta_{G^\prime(i), a}$ (industries), and {\tt{\small ind[[3]]}} is the $N\times L$ matrix $\delta_{G^{\prime\prime}(i), \alpha}$ (sectors), where the map $G$ is defined in (\ref{G.map}) (tickers to sub-industries), $G^\prime = GS$ (tickers to industries), and $G^{\prime\prime} = GSW$ (tickers to sectors), with the map $S$ (sub-industries to industries) defined in (\ref{S.map}), and the map $W$ (industries to sectors) defined in (\ref{W.map})); iii) {\tt{\small mkt.fac}}, where for {\tt{\small TRUE}} at the final step we have a single industry factor (``market"), while for {\tt{\small FALSE}} (default) the industry factors correspond to the least granular level in the industry classification (sectors for BICS); and iv) {\tt{\small rm.sing.tkr}}, where for {\tt{\small TRUE}} the tickers corresponding to the single-ticker ``clusters" at the most granular level in the industry classification (in the BICS case this would be the sub-industry level) are dropped altogether, while for {\tt{\small FALSE}} (default) the output universe is the same as the input universe. When the parameter {\tt{\small p}} is greater than 0, then the first $p$ principal components of the sample correlation matrix $\Psi_{ij}$ are appended (via {\tt{\small cbind()}}) to the $N\times K$ sub-industry factor covariance matrix $\Omega_{iA}$. Similarly, when the parameter {\tt{\small append.style = T}}, then the $N\times Y$ style factor matrix {\tt{\small style}} is appended to $\Omega_{iA}$. When both {\tt{\small p}} is greater than 0 and {\tt{\small append.style = T}}, then both the first $p$ principal components and the style factors are appended. When {\tt{\small p = 0}} and {\tt{\small append.style = F}}, there is another option, to wit, to have positive integer {\tt{\small k.style}}, in which case the code computes a heterotic CAPM model based on the first $k_{style}$ columns in {\tt{\small style}}. More precisely, for small sub-industries the code adjusts the number of style factors such that the resulting factor model for a given sub-industry is nonsingular (see Section \ref{sec.5}). The output is a list: {\tt{\small result\$spec.risk}} is the actual specific risk ${\overline \xi}_i$, {\tt{\small result\$fac.load}} is the actual factor loadings matrix ${\widetilde\Omega}_{i{\widetilde A}}$, {\tt{\small result\$fac.cov}} is the factor covariance matrix ${\overline \Phi}^*_{{\widetilde A}{\widetilde B}}$ (see Subsection \ref{sub.inv}), {\tt{\small result\$cov.mat}} is the factor model covariance matrix $\Gamma_{ij}$, and {\tt{\small result\$inv.cov}} is the matrix $\Gamma^{-1}_{ij}$ inverse to $\Gamma_{ij}$.

{}The function {\tt{\small qrm.style(prc, do.norm = T)}} computes the style factors ``prc", ``mom", ``hlv"and ``vol" of \cite{4F}. The input is: i) {\tt{\small prc}}, a list with daily price-volume data ({\tt{\small prc\$close}} is an $N\times d$ matrix with closing prices for the $d$ days in the time series (first column = most recent date), and similarly for open, high, low and volume); ii) if {\tt{\small do.norm = T}} the ``hlv" and ``vol" factors are normalized using the auxiliary function {\tt{\small qrm.normalize(x, center = mean(x), sdev = sd(x))}}, which conforms a vector {\tt{x}} to a normal distribution with the mean and standard deviation equal {\tt{\small center}} and {\tt{\small sdev}}, respectively; otherwise, said factors are unnormalized. In all backtests (see the main text), all tables and Figure 1 we have set {\tt{\small do.norm = F}}.

{}Finally, we give an alternative version of the function {\tt{\small qrm.cov.gen(ret, load, calc.inv = T)}} named {\tt{\small qrm.cov.gen.alt(ret, load, calc.inv = T)}}. The input and output are unchanged. The difference is in how the latter handles the ${\widetilde\Gamma}_{ii}\neq 1$ cases. The former uses the scale factors $\gamma_i$ as in Section \ref{sec.4} to simultaneously rescale the specific risk and the factor risk. The latter rescales only the factor risk in such a way that the specific variances defined as sample variances less diagonal elements of the factor risk are positive. The function {\tt{\small qrm.cov.gen.alt()}} calls another (self-explanatory) function {\tt{\small qrm.fr(fr, tv, low = (.1)\^{}2, high = (.9)\^{}2)}}, which rescales the factor risk. Generally, {\tt{\small qrm.cov.gen.alt()}} appears to underperform {\tt{\small qrm.cov.gen()}}. E.g., the ``prc" factor based heterotic CAPM model using {\tt{\small qrm.cov.gen.alt()}} backtests as follows (without any bounds): ROC 52.23\%, SR 13.85, CPS 2.84 (cf. Table \ref{table2}, row 3).
\\
\\
{\tt{\small
qrm.cov.gen <- function (ret, load, calc.inv = T)\\
\{\\
\indent print("Running qrm.cov.gen()...")\\
\\
\indent tr <- sqrt(apply(ret, 1, var))\\
\indent r1 <- ret / tr\\
\\
\indent reg <- lm(r1 $\sim$ -1 + load)\\
\indent z <- t(coef(reg))\\
\indent x <- residuals(reg)\\
\indent g <- var(z, z)\\
\indent x.f <- load \%*\% g \%*\% t(load)\\
\indent x.s <- apply(x, 1, var)\\
\indent tr1 <- sqrt(x.s + diag(x.f))\\
\indent tr <- tr / tr1\\
\indent sv <- x.s * tr\^{}2\\
\indent load <- load * tr\\
\indent cov.mat <- diag(sv) + t(x.f * tr) * tr\\
\\
\indent if(calc.inv)\\
\indent \{\\
\indent \indent v <- load / sv\\
\indent \indent d <- solve(g) + t(load) \%*\% v\\
\indent \indent inv <- diag(1 / sv) -  v \%*\% solve(d) \%*\% t(v)\\
\indent \}\\
\\\
\indent result <- new.env()\\
\indent result\$spec.risk <- sqrt(sv)\\
\indent result\$fac.load <- load\\
\indent result\$fac.cov <- g\\
\indent result\$cov.mat <- cov.mat\\
\indent result\$fac.ret <- t(z)\\
\indent result <- as.list(result)\\
\indent if(calc.inv)\\
\indent \indent result\$inv.cov <- inv\\
\indent return(result)\\
\}\\
\\
\noindent qrm.gen.het <- function (ret, ind, mkt.fac = F, rm.sing.tkr = F, p = 0,\\
\indent \indent \indent \indent \indent \indent \indent \indent \indent append.style = F, k.style = 0, style = 0)\\
\{\\
\indent print("Running qrm.gen.het()...")\\
\\
\indent null.elem <- function(z, prec = 1e-10)\\
\indent \{\\
\indent \indent bad <- sum(z == 0 | z < max(z) * prec)\\
\indent \indent return(bad)\\
\indent \}\\
\\
\indent s <- 0\\
\indent if(length(style) > 1)\\
\indent \indent s <- ncol(style)\\
\indent else\\
\indent \indent append.style <- F\\
\\
\indent if(append.style | p > 0)\\
\indent \indent k.style <- 0\\
\\
\indent if(mkt.fac)\\
\indent \indent ind[[length(ind) + 1]] <- matrix(1, nrow(ind[[1]]), 1)\\
\\
\indent if(rm.sing.tkr)\\
\indent \{\\
\indent \indent bad <- colSums(ind[[1]]) == 1\\
\indent \indent ind[[1]] <- ind[[1]][, !bad]\\
\indent \indent bad <- rowSums(ind[[1]]) == 0\\
\indent \indent for(lvl in 1:length(ind))\\
\indent \indent \indent ind[[lvl]] <- ind[[lvl]][!bad, ]\\
\indent \indent ret <- ret[!bad, ]\\
\\
\indent \indent if(length(style) > 1)\\
\indent \indent \indent if(ncol(style) > 1)\\
\indent \indent \indent \indent style <- style[!bad, ]\\
\indent \indent \indent else\\
\indent \indent \indent \indent style <- matrix(style[!bad, ], length(style[!bad, ]), 1)\\
\indent \}\\
\\
\indent flm <- ind\\
\indent fac.ret <- list()\\
\indent spec.risk <- list()\\
\indent fac.cov <- list()\\
\indent fac.ret[[1]] <- ret\\
\\
\indent if(k.style > 0)\\
\indent \{\\
\indent \indent x <- y <- rep(NA, nrow(ind[[1]]))\\
\indent \indent for(a in 1:ncol(ind[[1]]))\\
\indent \indent \{\\
\indent \indent \indent take <- ind[[1]][, a] > 0\\
\indent \indent \indent k <- sum(take)\\
\indent \indent \indent for(j in 1:ncol(style))\\
\indent \indent \indent \indent if(sum(style[take, j] == 0) == k)\\
\indent \indent \indent \indent \indent style[take, j] <- 1\\
\\
\indent \indent \indent chk.mat <- k > 1\\
\indent \indent \indent if(k < k.style)\\
\indent \indent \indent \indent k <- 1\\
\indent \indent \indent if(k > 1)\\
\indent \indent \indent \indent k <- min(k - 1, ncol(style))\\
\indent \indent \indent if(chk.mat \& k > 1)\\
\indent \indent \indent \indent if(null.elem(eigen(t(style[take, 1:k]) \%*\% \\
\indent \indent \indent \indent \indent \indent \indent style[take, 1:k])\$values)> 0)\\
\indent \indent \indent \indent \indent k <- 1\\
\\
\indent \indent \indent while(chk.mat)\\
\indent \indent \indent \{\\
\indent \indent \indent \indent z <- residuals(lm(fac.ret[[1]][take, ] $\sim$ -1 +\\
\indent \indent \indent \indent \indent \indent style[take, 1:k]))\\
\indent \indent \indent \indent z <- apply(z, 1, sd)\\
\indent \indent \indent \indent bad <- null.elem(z)\\
\indent \indent \indent \indent if(chk.mat <- bad > 0)\\
\indent \indent \indent \indent \indent if(bad < k)\\
\indent \indent \indent \indent \indent \indent k <- k - bad\\
\indent \indent \indent \indent \indent else\\
\indent \indent \indent \indent \indent \{\\
\indent \indent \indent \indent \indent \indent k <- 1\\
\indent \indent \indent \indent \indent \indent style[take, 1] <- 1\\
\indent \indent \indent \indent \indent \indent chk.mat <- F\\
\indent \indent \indent \indent \indent \}\\
\indent \indent \indent \}\\
\indent \indent \indent x <- cbind(x, matrix(ind[[1]][, a] * style[, 1:k], \\
\indent \indent \indent \indent \indent nrow(ind[[1]]), k))\\
\indent \indent \indent y <- cbind(y, matrix(ind[[1]][, a], nrow(ind[[1]]), k))\\
\indent \indent \}\\
\indent \indent flm[[1]] <- x[, -1]\\
\indent \indent ind[[1]] <- y[, -1]\\
\indent \}\\
\\
\indent calc.load <- function(load, load1)\\
\indent \{\\
\indent \indent x <- colSums(load1)\\
\indent \indent load <- (t(load1) \%*\% load) / x\\
\indent \indent return(load)\\
\indent \}\\
\\
\indent for(lvl in 1:length(ind))\\
\indent \{\\
\indent \indent if(lvl > 1)\\
\indent \indent \indent flm[[lvl]] <- calc.load(ind[[lvl]], ind[[lvl - 1]])\\
\\
\indent \indent if(k.style == 0 | lvl > 1)\\
\indent \indent \indent for(a in 1:ncol(flm[[lvl]]))\\
\indent \indent \indent \{\\
\indent \indent \indent \indent take <- as.logical(flm[[lvl]][, a])\\
\\
\indent \indent \indent \indent if(lvl == 1)\\
\indent \indent \indent \indent \indent k <- nrow(fac.ret[[lvl]])\\
\indent \indent \indent \indent else\\
\indent \indent \indent \indent \indent k <- ncol(ind[[lvl - 1]])\\
\\
\indent \indent \indent \indent x <- matrix(fac.ret[[lvl]][1:k, ][take, ], sum(take),\\
\indent \indent \indent \indent \indent \indent ncol(fac.ret[[lvl]]))\\
\indent \indent \indent \indent y <- eigen(cor(t(x)))\$vectors\\
\indent \indent \indent \indent y1 <- y[, 1]\\
\indent \indent \indent \indent flm[[lvl]][take, a] <- y1 * flm[[lvl]][take, a]\\
\indent \indent \indent \}\\
\\
\indent \indent if(p > 0 | append.style)\\
\indent \indent \indent if(lvl == 1)\\
\indent \indent \indent \{\\
\indent \indent \indent \indent s <- 0\\
\indent \indent \indent \indent if(length(style) > 1)\\
\indent \indent \indent \indent \indent s <- ncol(style)\\
\indent \indent \indent \indent tmp <- matrix(0, nrow(flm[[lvl]]), p + s)\\
\indent \indent \indent \indent if(p > 0)\\
\indent \indent \indent \indent \indent tmp[, 1:p] <- eigen(cor(t(ret)))\$vectors[, 1:p]\\
\indent \indent \indent \indent if(s > 0)\\
\indent \indent \indent \indent \indent tmp[, (p+1):(p+s)] <- style\\
\indent \indent \indent \indent flm[[lvl]] <- cbind(flm[[lvl]], tmp)\\
\indent \indent \indent \indent p <- p + s\\
\indent \indent \indent \}\\
\indent \indent \indent else\\
\indent \indent \indent \{\\
\indent \indent \indent \indent tmp <- matrix(0, nrow(flm[[lvl]]) + p, ncol(flm[[lvl]]) + p)\\
\indent \indent \indent \indent tmp[1:nrow(flm[[lvl]]), 1:ncol(flm[[lvl]])] <- flm[[lvl]]\\
\indent \indent \indent \indent tmp[(nrow(flm[[lvl]]) + 1):(nrow(flm[[lvl]]) + p),\\
\indent \indent \indent \indent\indent (ncol(flm[[lvl]]) + 1):(ncol(flm[[lvl]]) + p)]\\
\indent \indent \indent \indent \indent \indent <- diag(1, p)\\
\indent \indent \indent \indent flm[[lvl]] <- tmp\\
\indent \indent \indent \}\\
\\
\indent \indent res <- qrm.cov.gen(fac.ret[[lvl]], flm[[lvl]], calc.inv = F)\\
\indent \indent spec.risk[[lvl]] <- res\$spec.risk\\
\indent \indent fac.cov[[lvl]] <- res\$fac.cov\\
\indent \indent flm[[lvl]] <- res\$fac.load\\
\indent \indent fac.ret[[lvl + 1]] <- res\$fac.ret\\
\indent \}\\
\\
\indent for(lvl in length(ind):1)\\
\indent \indent if(lvl > 1)\\
\indent \indent \indent fac.cov[[lvl - 1]] <- diag(spec.risk[[lvl]]\^{}2) +\\
\indent \indent \indent \indent \indent \indent flm[[lvl]] \%*\% fac.cov[[lvl]] \%*\% t(flm[[lvl]])\\
\indent \indent else\\
\indent \indent \{\\
\indent \indent \indent spec.risk <- spec.risk[[1]]\\
\indent \indent \indent fac.cov <- fac.cov[[1]]\\
\indent \indent \indent flm <- flm[[1]]\\
\indent \indent \indent mod.mat <- diag(spec.risk\^{}2) + flm \%*\% fac.cov \%*\% t(flm)\\
\indent \indent \}\\
\\
\indent sv <- spec.risk\^{}2\\
\\
\indent if(!rm.sing.tkr)\\
\indent \{\\
\indent \indent k <- ncol(ind[[1]])\\
\indent \indent sv1 <- colSums((t(flm[, 1:k]))\^{}2 * diag(fac.cov[1:k, 1:k]))\\
\indent \indent take <- colSums(ind[[1]]) == 1\\
\indent \indent x <- diag(fac.cov)\\
\indent \indent y <- x[1:k]\\
\indent \indent y[take] <- 0\\
\indent \indent x[1:k] <- y\\
\indent \indent diag(fac.cov) <- x\\
\indent \indent take <- rowSums(ind[[1]][, take]) == 1\\
\indent \indent sv[take] <- sv1[take]\\
\indent \indent spec.risk <- sqrt(sv)\\
\indent \}\\
\\
\indent v <- flm / sv\\
\indent d <- solve(fac.cov) + t(flm) \%*\% v\\
\indent inv <- diag(1 / sv) -  v \%*\% solve(d) \%*\% t(v)\\
\\
\indent result <- new.env()\\
\indent result\$spec.risk <- spec.risk\\
\indent result\$fac.load <- flm\\
\indent result\$fac.cov <- fac.cov\\
\indent result\$cov.mat <- mod.mat\\
\indent result\$inv.cov <- inv\\
\indent result <- as.list(result)\\
\indent return(result)\\
\}\\
\\
\noindent qrm.normalize <- function(x, center = mean(x), sdev = sd(x))\\
\indent qnorm(ppoints(x)[sort.list(sort.list(x), method = "radix")],\\
\indent \indent \indent center, sdev)\\
\\
\noindent qrm.style <- function (prc, do.norm = T)\\
\{\\
\indent open <- prc\$open\\
\indent close <- prc\$close\\
\indent high <- prc\$high\\
\indent low <- prc\$low\\
\indent vol <- prc\$vol\\
\\
\indent \#\#\# prc factor\\
\indent st1 <- log(close[, 1])\\
\indent \#\#\# mom factor\\
\indent st2 <- log(close[, 1]/open[, 1])\\
\indent \#\#\# hlv factor\\
\indent st3 <- .5 * log(apply(((high - low)/close)\^{}2, 1, mean))\\
\indent if(do.norm)\\
\indent \indent st3 <- qrm.normalize(st3)\\
\indent \#\#\# vol factor\\
\indent st4 <- log(apply(vol, 1, mean))\\
\indent if(do.norm)\\
\indent \indent st4 <- qrm.normalize(st4)\\
\\
\indent prc\$style <- cbind(st1, st2, st3, st4)\\
\}\\
\\
\noindent qrm.fr <- function(fr, tv, low = (.1)\^{}2, high = (.9)\^{}2)\\
\{\\
\indent y <- log(fr)\\
\indent y <- y  - .5 * log(tv)\\
\indent y <- qrm.normalize(y, median(y), mad(y))\\
\indent min.y <- .5 * log(low)\\
\indent max.y <- .5 * log(high)\\
\indent center <- median(y)\\
\indent \indent sdev <- mad(y)\\
\indent \indent x <- (y - center) / sdev\\
\indent min.x <- max((min.y - center) / sdev, min(x))\\
\indent max.x <- min((max.y - center) / sdev, max(x))\\
\indent if(min.x < 0)\\
\indent \indent x[x < 0] <- min.x * (1 - exp(x[x < 0]))\\
\indent else\\
\indent \indent x[x < min.x] <- min.x\\
\indent if(max.x > 0)\\
\indent \indent x[x > 0] <- max.x * (1 - exp(-x[x > 0]))\\
\indent else\\
\indent \indent x[x > max.x] <- max.x\\
\indent y <- center + x * sdev\\
\indent y <- exp(y) * sqrt(tv)\\
\indent return(y)\\
\}\\
\\
qrm.cov.gen.alt <- function (ret, load, ind, calc.inv = T)\\
\{\\
\indent print("Running qrm.cov.gen.alt()...")\\
\\
\indent tr <- sqrt(tv <- apply(ret, 1, var))\\
\indent r1 <- ret / tr\\
\\
\indent reg <- lm(r1 $\sim$ -1 + load)\\
\indent z <- t(coef(reg))\\
\indent g <- var(z, z)\\
\indent load <- tr * load\\
\indent x.f <- load \%*\% g \%*\% t(load)\\
\indent fr <- sqrt(diag(x.f))\\
\indent y <- qrm.fr(fr, tv)\\
\indent take <- fr > 0\\
\indent y[!take] <- 0\\
\indent sv <- tv - y\^{}2\\
\indent y[take] <- y[take] / fr[take]\\
\indent load <- load * y\\
\indent cov.mat <- diag(sv) + t(x.f * y) * y\\
\\
\indent if(calc.inv)\\
\indent \{\\
\indent \indent v <- load / sv\\
\indent \indent d <- solve(g) + t(load) \%*\% v\\
\indent \indent inv <- diag(1 / sv) -  v \%*\% solve(d) \%*\% t(v)\\
\indent \}\\
\\
\indent result <- new.env()\\
\indent result\$spec.risk <- sqrt(sv)\\
\indent result\$fac.load <- load\\
\indent result\$fac.cov <- g\\
\indent result\$cov.mat <- cov.mat\\
\indent result\$fac.ret <- t(z)\\
\indent result <- as.list(result)\\
\indent if(calc.inv)\\
\indent \indent result\$inv.cov <- inv\\
\indent return(result)\\
\}
}}

\section{DISCLAIMERS}\label{app.B}

{}Wherever the context so requires, the masculine gender includes the feminine and/or neuter, and the singular form includes the plural and {\em vice versa}. The author of this paper (``Author") and his affiliates including without limitation Quantigic$^\circledR$ Solutions LLC (``Author's Affiliates" or ``his Affiliates") make no implied or express warranties or any other representations whatsoever, including without limitation implied warranties of merchantability and fitness for a particular purpose, in connection with or with regard to the content of this paper including without limitation any code or algorithms contained herein (``Content").

{}The reader may use the Content solely at his/her/its own risk and the reader shall have no claims whatsoever against the Author or his Affiliates and the Author and his Affiliates shall have no liability whatsoever to the reader or any third party whatsoever for any loss, expense, opportunity cost, damages or any other adverse effects whatsoever relating to or arising from the use of the Content by the reader including without any limitation whatsoever: any direct, indirect, incidental, special, consequential or any other damages incurred by the reader, however caused and under any theory of liability; any loss of profit (whether incurred directly or indirectly), any loss of goodwill or reputation, any loss of data suffered, cost of procurement of substitute goods or services, or any other tangible or intangible loss; any reliance placed by the reader on the completeness, accuracy or existence of the Content or any other effect of using the Content; and any and all other adversities or negative effects the reader might encounter in using the Content irrespective of whether the Author or his Affiliates is or are or should have been aware of such adversities or negative effects.

{}The R code included in Appendix \ref{app.A} hereof is part of the copyrighted R code of Quantigic$^\circledR$ Solutions LLC and is provided herein with the express permission of Quantigic$^\circledR$ Solutions LLC. The copyright owner retains all rights, title and interest in and to its copyrighted source code included in Appendix \ref{app.A} hereof and any and all copyrights therefor.



\begin{table}[ht]
\caption{Simulation results for the optimized alphas discussed in Subsection \ref{sub.opt} using the heterotic risk model with various style and/or principal component risk factors appended to the industry factors, without any bounds on the dollar holdings. All quantities are rounded to 2 digits. The result in the third line is the same as in \cite{Het} with the slight difference in CPS due to rounding down employed therein. See Subsection \ref{sub.opt} for details.} 
\begin{tabular}{l l l l} 
\\
\hline\hline 
Risk Model & ROC & SR & CPS\\[0.5ex] 
\hline 
Heterotic, BICS sectors & 50.88\% & 13.04 & 2.41\\
Heterotic, BICS industries & 54.26\% & 14.45 & 2.56\\
Heterotic, BICS sub-industries & 55.90\% & 15.41 & 2.68\\
Heterotic, BICS sub-industries + prc & 55.92\% & 15.47 & 2.68\\
Heterotic, BICS sub-industries + hlv & 55.88\% & 15.50 & 2.68\\
Heterotic, BICS sub-industries + vol & 56.02\% & 15.46 & 2.68\\
Heterotic, BICS sub-industries + mom & 56.09\% & 14.71 & 2.65\\
Heterotic, BICS sub-industries + prc + hlv + vol & 55.85\% & 15.80 & 2.67\\
Heterotic, BICS sub-industries + 1st prin.comp & 55.92\% & 15.47 & 2.67\\
Heterotic, BICS sub-industries + 1st 10 prin.comps & 55.19\% & 15.15 & 2.66\\
Heterotic, BICS sub-industries + 1st prin.com + prc & 55.97\% & 15.55 & 2.68\\[1ex] 
\hline 
\end{tabular}
\label{table1} 
\end{table}

\begin{table}[ht]
\caption{Simulation results for the optimized alphas using the heterotic CAPM risk model based on various style factors, without any bounds on the dollar holdings. All quantities are rounded to 2 digits. The result in the first line is the same as in \cite{Het} with the slight difference in CPS due to rounding down employed therein. See Subsection \ref{sub.capm.q} for details.} 
\begin{tabular}{l l l l} 
\\
\hline\hline 
Risk Model & ROC & SR & CPS\\[0.5ex] 
\hline 
Heterotic, BICS sub-industries & 55.90\% & 15.41 & 2.68\\
Heterotic CAPM, int & 55.99\% & 15.04 & 2.60\\
Heterotic CAPM, prc & 55.06\% & 15.24 &  2.99\\
Heterotic CAPM, hlv & 55.21\% & 16.07 & 2.66\\
Heterotic CAPM, vol & 54.82\% & 14.59 & 2.50\\
Heterotic CAPM, mom & 45.86\% & 9.92 & 1.98\\[1ex] 
\hline 
\end{tabular}
\label{table2} 
\end{table}

\begin{table}[ht]
\caption{Simulation results for the heterotic CAPM model based on the ``prc" style factor for various values of the weight $q$ defined in (\ref{capm.q}) in Subsection \ref{sub.capm.q}. ROC (return-on-capital) decreases with increasing $q$, SR (Sharpe ratio) peaks around $q=0.4$, while CPS (cents-per-share) peaks around $q=0.6$.} 
\begin{tabular}{l l l l} 
\\
\hline\hline 
$q$ & ROC & SR & CPS\\[0.5ex] 
\hline 
0 & 55.99\% & 15.04 & 2.60\\
0.1 & 55.99\% & 15.11 & 2.65\\
0.2 & 55.94\% & 15.19 & 2.72\\
0.3 & 55.83\% & 15.25 & 2.80\\
0.4 & 55.58\% & 15.29 & 2.89\\
0.5 & 55.06\% & 15.24 & 2.99\\
0.6 & 54.01\% & 14.97 & 3.03\\
0.7 & 51.99\% & 14.21 & 2.96\\
0.8 & 48.84\% & 12.81 & 2.73\\
0.9 & 45.36\% & 11.04 & 2.31\\
1.0 & 42.44\% & 9.91 & 1.86\\[1ex] 
\hline 
\end{tabular}
\label{table3} 
\end{table}

\begin{table}[ht]
\caption{Simulation results for the heterotic CAPM model based on the ``hlv" style factor for various values of the weight $q$ defined in (\ref{capm.q}) in Subsection \ref{sub.capm.q}. ROC (return-on-capital) decreases with increasing $q$, while both SR (Sharpe ratio) and CPS (cents-per-share) peak around $q=0.7$.} 
\begin{tabular}{l l l l} 
\\
\hline\hline 
$q$ & ROC & SR & CPS\\[0.5ex] 
\hline 
0 & 55.99\% & 15.04 & 2.60\\
0.1 & 55.91\% & 15.16 & 2.61\\
0.2 & 55.81\% & 15.32 & 2.62\\
0.3 & 55.68\% & 15.51 & 2.63\\
0.4 & 55.49\% & 15.76 & 2.64\\
0.5 & 55.21\% & 16.07 &  2.66\\
0.6 & 54.73\% & 16.46 & 2.68\\
0.7 & 53.76\% & 16.82 & 2.70\\
0.8 & 51.23\% & 16.43 & 2.64\\
0.9 & 46.31\% & 13.88 & 2.39\\
1.0 & 43.31\% & 11.02 & 1.99\\[1ex] 
\hline 
\end{tabular}
\label{table4} 
\end{table}

\begin{table}[ht]
\caption{Simulation results for the vanilla heterotic and heterotic CAPM models with the bounds (\ref{liq}). The result in the first line is the same as in \cite{Het} with the slight difference in CPS due to rounding down employed therein.} 
\begin{tabular}{l l l l} 
\\
\hline\hline 
Risk Model & ROC & SR & CPS\\[0.5ex] 
\hline 
Heterotic, BICS sub-industries & 49.00\% & 19.23 & 2.37\\
Heterotic CAPM, int & 49.24\% & 19.04 & 2.31\\
Heterotic CAPM, prc & 48.47\% & 18.89 & 2.64\\[1ex] 
\hline 
\end{tabular}
\label{table5} 
\end{table}

\begin{table}[ht]
\caption{Simulation results for the heterotic model with the first principal component for each sub-industry replaced by the $k_{pc}$-th principal component (or the number of tickers in said sub-industry if this number is smaller than $k_{pc}$; see Subsection \ref{sub.kappa}). The first row ($k_{pc} = 1$) corresponds to the original heterotic risk model.} 
\begin{tabular}{l l l l} 
\\
\hline\hline 
$k_{pc}$ & ROC & SR & CPS\\[0.5ex] 
\hline 
1 & 55.90\% & 15.41 & 2.68\\
2 & 37.97\% & 8.22 & 1.66\\
3 & 38.25\% & 8.19 & 1.68\\
4 & 38.21\% & 8.09 & 1.68\\
5 & 38.31\% & 8.11 & 1.68\\[1ex] 
\hline 
\end{tabular}
\label{table6} 
\end{table}

\newpage

\begin{figure}[ht]
\centerline{\epsfxsize 4.truein \epsfysize 4.truein\epsfbox{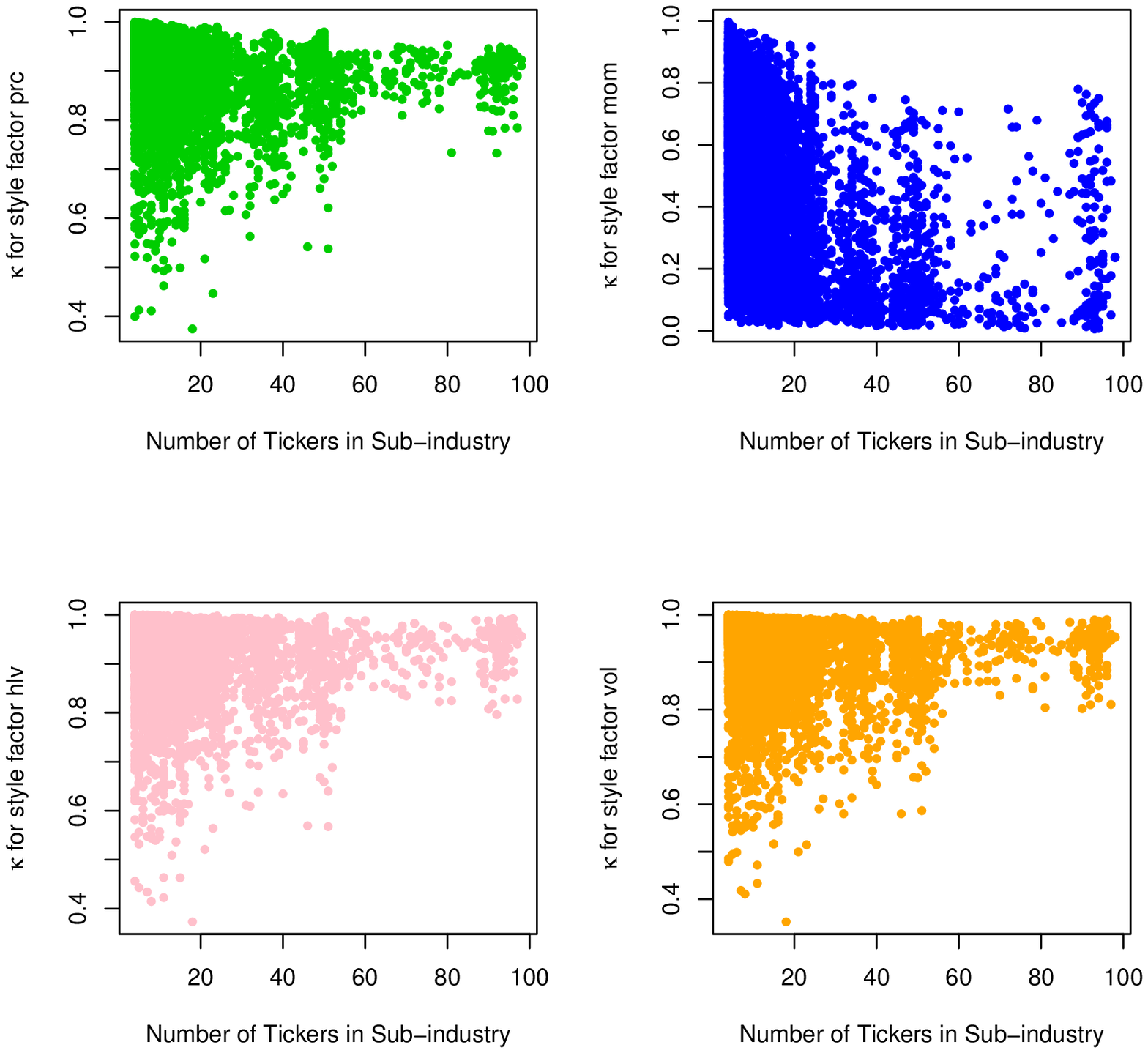}}
\noindent{\small {Figure 1. The values of the coefficient $\kappa(A)$ defined in (\ref{kappa.sub}) vs. the number $N(A)$ of tickers in sub-industries for the style factors ``prc", ``mom", ``hlv" and ``vol". See Subsection \ref{sub.kappa} for details.}}
\end{figure}


\begin{thebibliography}{99}

\makeatletter
\def\@biblabel#1{}
\makeatother

\bibitem[Acharya and Pedersen, 2005]{Q1} Acharya, V.V. and Pedersen, L.H. (2005)
Asset pricing with liquidity risk.
{\em Journal of Financial Economics} 77(2): 375-410.

\bibitem[Ang {\em et al}, 2006]{Q2} Ang, A., Hodrick, R., Xing, Y. and Zhang, X. (2006)
The Cross-Section of Volatility and Expected Returns.
{\em Journal of Finance} 61(1): 259-299.

\bibitem[Anson, 2013/14]{Q3} Anson, M. (2013/14) Performance Measurement in Private Equity: Another Look at the Lagged Beta Effect.
{\em The Journal of Private Equity} 17(1): 29-44.

\bibitem[Asness, 1995]{Q4} Asness, C.S. (1995)
The Power of Past Stock Returns to Explain Future Stock Returns.
Goldman Sachs Asset Management. Working paper.

\bibitem[Asness and Stevens, 1995]{Q5} Asness, C. and Stevens, R. (1995)
Intra- and Inter-Industry Variation in the Cross-Section of Expected Stock Returns.
Goldman Sachs Asset Management. Working paper.

\bibitem[Asness {\em et al}, 2001]{Q6} Asness, C., Krail, R.J. and Liew, J.M. (2001)
Do Hedge Funds Hedge?
{\em The Journal of Portfolio Management} 28(1): 6-19.

\bibitem[Bai, 2003]{Q7} Bai, J. (2003)
Inferential theory for factor models of large dimension.
{\em Econometrica} 71(11): 135-171.

\bibitem[Bai and Li, 2012]{Q8} Bai, J. and Li, K. (2012)
Statistical Analysis of Factor Models of High Dimension.
{\em The Annals of Statistics} 40(1): 436-465.

\bibitem[Bai and Ng, 2002]{Q9} Bai, J. and Ng, S. (2002)
Determining the number of factors in approximate factor models.
{\em Econometrica} 70(1): 191-221.

\bibitem[Bansal and Viswanathan, 1993]{Q10} Bansal, R. and Viswanathan, S. (1993)
No Arbitrage and Arbitrage Pricing: A New Approach.
{\em The Journal of Finance} 48(4): 1231-1262.

\bibitem[Banz, 1981]{Q11} Banz, R. (1981)
The relationship between return and market value of common stocks.
{\em Journal of Financial Economics} 9(1): 3-18.

\bibitem[Basu, 1977]{Q12} Basu, S. (1977)
The investment performance of common stocks in relation to their price to earnings ratios: A test of the efficient market hypothesis.
{\em Journal of Finance} 32(3): 663-682.

\bibitem[Black, 1972]{Q13} Black, F. (1972)
Capital market equilibrium with restricted borrowing.
{\em Journal of Business} 45(3): 444-455.

\bibitem[Black {\em et al}, 1972]{Q14} Black, F., Jensen, M. and Scholes, M. (1972)
The capital asset pricing model: Some empirical tests.
In: Jensen, M. (ed.) {\em Studies in the Theory of Capital Markets}. New York, NY: Praeger Publishers, pp. 79-121.

\bibitem[Blume and Friend, 1973]{Q15} Blume, O. and Friend, L. (1973) A new look at the capital asset pricing model.
{\em Journal of Finance} 28(1): 19-33.

\bibitem[Brandt {\em et al}, 2010]{Q16} Brandt, M.W., Brav, A., Graham, J.R. and Kumar, A. (2010)
The idiosyncratic volatility puzzle: Time trend or speculative episodes?
{\em Review of Financial Studies} 23(2): 863-899.

\bibitem[Briner and Connor, 2008]{Q17} Briner, B. and Connor, G. (2008)
How much structure is best? A comparison of market model, factor model and unstructured equity covariance matrices.
{\em Journal of Risk} 10(4), 3-30.

\bibitem[Burmeister and Wall, 1986]{Q18} Burmeister, E. and Wall, K.D. (1986)
The arbitrage pricing theory and macroeconomic factor measures.
{\em Financial Review} 21(1): 1-20.

\bibitem[Campbell, 1987]{Q19} Campbell, J. (1987)
Stock returns and the term structure.
{\em Journal of Financial Economics} 18(2): 373-399.

\bibitem[Campbell {\em et al}, 2001]{Q20} Campbell, J.Y., Lettau, M., Malkiel, B.G. and Xu, Y. (2001)
Have individual stocks become more volatile? An empirical exploration of idiosyncratic risk.
{\em Journal of Finance} 56(1): 1-43.

\bibitem[Campbell and Shiller, 1988]{Q21} Campbell, J. and Shiller, R. (1988)
The dividend-price ratio and expectations of future dividends and discount factors.
{\em Review of Financial Studies} 1(3): 195-227.

\bibitem[Carhart, 1997]{Q22} Carhart, M.M. (1997)
Persistence in mutual fund performance.
{\em Journal of Finance} 52(1): 57-82.

\bibitem[Chamberlain and Rothschild, 1983]{Q23} Chamberlain, G. and Rothschild, M. (1983)
Arbitrage, Factor Structure, and Mean-Variance Analysis on Large Asset Markets.
{\em Econometrica} 51(5): 1281-1304.

\bibitem[Chan {\em et al}, 1985]{Q24} Chan, K.C., Chen, N. and Hsieh, D. (1985)
An Exploratory Investigation of the Firm Size Effect.
{\em Journal of Financial Economics} 14(3): 451-471.

\bibitem[\,1990]{Q25} Chen, N., Grundy, B. and Stambaugh, R.F. (1990)
Changing Risk, Changing Risk Premiums, and Dividend Yield Effects.
{\em The Journal of Business} 63(1): 51-70.

\bibitem[Chen {\em et al}, 1986]{Q26} Chen, N., Roll, R. and Ross, S. (1986)
Economic forces and the stock market.
{\em Journal of Business} 59(3): 383-403.

\bibitem[Chicheportiche and Bouchaud, 2015]{Q27} Chicheportiche, R. and Bouchaud, J.-P. (2015)
A nested factor model for non-linear dependencies in stock returns.
{\em Quantitative Finance} 15(11): 1789-1804.

\bibitem[Cochrane, 2001]{Q28} Cochrane, J.H. (2001)
{\em Asset Pricing.}
Princeton, NJ: Princeton University Press.

\bibitem[Connor, 1984]{Q29} Connor, G. (1984)
A unified beta pricing theory.
{\em Journal of Economic Theory} 34(1): 13-31.

\bibitem[\,1995]{Q30} Connor, G. (1995)
The Three Types of Factor Models: A Comparison of Their Explanatory Power.
{\em Financial Analysts Journal} 51(3): 42-46.

\bibitem[Connor and Korajczyk, 1988]{Q31} Connor, G. and Korajczyk, R. (1988)
Risk and return in an equilibrium APT: Application of a new test methodology.
{\em Journal of Financial Economics} 21(2): 255-289.

\bibitem[\,1989]{Q32} Connor, G. and Korajczyk, R. (1989)
An intertemporal beta pricing model.
{\em Review of Financial Studies} 2(3): 373-392.

\bibitem[\,2010]{Q33} Connor, G. and Korajczyk, R. (2010) Factor Models in Portfolio and Asset Pricing Theory.
In: Guerard Jr, J.B. (ed.) {\em Handbook of Portfolio Construction: Contemporary Applications of Markowitz Techniques}.
New York, NY: Springer, pp. 401-418.

\bibitem[Daniel and Titman, 1997]{Q34} Daniel, K. and Titman, S. (1997)
Evidence on the Characteristics of Cross Sectional Variation in Stock Returns.
{\em Journal of Finance} 52(1): 1-33.

\bibitem[DeBondt and Thaler, 1985]{Q35} DeBondt, W. and Thaler, R. (1985) Does the stock market overreact?
{\em Journal of Finance} 40(3): 739-805.

\bibitem[Dhrymes {\em et al}, 1984]{Q36} Dhrymes, P.J., Friend, I. and Gultekin, N.B. (1984)
A Critical Reexamination of the Empirical Evidence on the Arbitrage Pricing Theory.
{\em The Journal of Finance} 39(2): 323-346.

\bibitem[Ederington and Guan, 2002]{ImpliedVol} Ederington, L. and Guan, W. (2002)
Is implied volatility an informationally efficient and effective predictor of future volatility?
{\em The Journal of Risk} 4(3): 29-46.

\bibitem[Fama and French, 1992]{Q37} Fama, E. and French, K. (1992)
The cross-section of expected stock returns.
{\em Journal of Finance} 47(2): 427-465.

\bibitem[\,1993]{Q38} Fama, E.F. and French, K.R. (1993)
Common risk factors in the returns on stocks and bonds.
{\em J. Financ. Econ.} 33(1): 3-56.

\bibitem[\,1996]{Q39} Fama, E. and French, K. (1996)
Multifactor explanations for asset pricing anomalies.
{\em Journal of Finance} 51(1): 55-94.

\bibitem[\,2015]{Q40} Fama, E. and French, K. (2015)
A Five-Factor Asset Pricing Model.
{\em Journal of Financial Economics} (forthcoming), DOI: 10.1016/j.jfineco.2014.10.010.

\bibitem[Fama and McBeth, 1973]{Q41} Fama, E.F. and MacBeth, J.D. (1973)
Risk, Return and Equilibrium: Empirical Tests.
{\em Journal of Political Economy} 81(3): 607-636.

\bibitem[Ferson and Harvey, 1991]{Q42} Ferson, W. and Harvey, C. (1991)
The variation in economic risk premiums.
{\em Journal of Political Economy} 99(2): 385-415.

\bibitem[\,1999]{Q43} Ferson, W. and Harvey, C. (1999)
Conditioning variables and the cross section of stock returns.
{\em Journal of Finance} 54(4): 1325-1360.

\bibitem[Forni {\em et al}, 2000]{Q44} Forni, M., Hallin, M., Lippi, M. and Reichlin, L. (2000)
The generalized dynamic factor model: Identification and estimation.
{\em The Review of Economics and Statistics} 82(4): 540-554.

\bibitem[\,2005]{Q45} Forni, M., Hallin, M., Lippi, M. and Reichlin, L. (2005)
The generalized dynamic factor model: One-sided estimation and forecasting.
{\em Journal of the American Statistical Association} 100(471): 830-840.

\bibitem[Forni and Lippi 2001]{Q46} Forni, M. and Lippi, M. (2001)
The generalized dynamic factor model: Representation theory.
{\em Econometric Theory} 17(6): 1113-1141.

\bibitem[Goyal {\em et al}, 2008]{Q47} Goyal, A., Perignon, C. and Villa, C. (2008)
How common are common return factors across the NYSE and Nasdaq?
{\em Journal of Financial Economics} 90(3): 252-271.

\bibitem[Goyal and Santa-Clara, 2003]{Q48} Goyal, A. and Santa-Clara, P. (2003)
Idiosyncratic risk matters!
{\em Journal of Finance} 58(3): 975-1007.

\bibitem[Grinold and Kahn, 2000]{Q49} Grinold, R.C. and Kahn, R.N. (2000)
{\em Active Portfolio Management.} New York, NY: McGraw-Hill.

\bibitem[Hall {\em et al}, 2002]{Q50} Hall, A.D., Hwang, S. and Satchell, S.E. (2002)
Using bayesian variable selection methods to choose style factors in global stock return models.
{\em Journal of Banking and Finance} 26(12): 2301-2325.

\bibitem[Haugen, 1995]{Q51} Haugen, R.A. (1995)
{\em The New Finance: The Case Against Efficient Markets.}
Upper Saddle River, NJ: Prentice Hall.

\bibitem[Heaton and Lucas, 1999]{Q52} Heaton, J. and Lucas, D.J. (1999)
Stock Prices and Fundamentals.
{\em NBER Macroeconomics Annual} 14(1): 213-242.

\bibitem[Heston and Rouwenhorst, 1994]{Q53} Heston, S.L. and Rouwenhorst, K.G. (1994)
Does Industrial Structure Explain the Benefits of International Diversification?
{\em Journal of Financial Economics} 36(1): 3-27.

\bibitem[Hong, Torous and Valkanov, 2007]{ind.lead} Hong, H., Torous, W. and Valkanov, R. (2007)
{\em Journal of Financial Economics} 83(2): 367-396.

\bibitem[Jagannathan and Wang, 1996]{Q54} Jagannathan, R. and Wang, Z. (1996)
The conditional CAPM and the cross-section of expected returns.
{\em Journal of Finance} 51(1): 3-53.

\bibitem[Jegadeesh and Titman, 1993]{Q55} Jegadeesh, N. and Titman, S. (1993)
Returns to buying winners and selling losers: Implications for stock market efficiency.
{\em Journal of Finance} 48(1): 65-91.

\bibitem[\,2001]{Q56} Jegadeesh, N. and Titman, S. (2001)
Profitability of Momentum Strategies: An Evaluation of Alternative Explanations.
{\em Journal of Finance} 56(2): 699-720.

\bibitem[Kakushadze, 2015a]{MeanRev} Kakushadze, Z. (2015a)
Mean-Reversion and Optimization.
{\em Journal of Asset Management} 16(1): 14-40.\\
Available online: http://ssrn.com/abstract=2478345.

\bibitem[Kakushadze, 2015b]{4F} Kakushadze, Z. (2015b)
4-Factor Model for Overnight Returns. {\em Wilmott Magazine} 2015(79): 56-62. Available online: http://ssrn.com/abstract=251187.

\bibitem[Kakushadze, 2015c]{RusDoll} Kakushadze, Z. (2015c)
Russian-Doll Risk Models.
{\em Journal of Asset Management} 16(3): 170-185. Available online: http://ssrn.com/abstract=2538123.

\bibitem[Kakushadze, 2015d]{Het} Kakushadze, Z. (2015d)
Heterotic Risk Models. {\em Wilmott Magazine} 2015(80): 40-55. Available online: http://ssrn.com/abstract=2600798.

\bibitem[Kakushadze, 2015e] {KLT} Kakushadze, Z. (2015e)
101 Formulaic Alphas. {\em Wilmott Magazine} (forthcoming).
Available online: http://ssrn.com/abstract=2701346.

\bibitem[Kakushadze and Liew, 2015]{CustomRM} Kakushadze, Z. and Liew, J.K.-S. (2015)
Custom v. Standardized Risk Models.
{\em Risks} 3(2): 112-138. Available online: http://ssrn.com/abstract=2493379.

\bibitem[King, 1966]{Q62} King, B.F. (1966)
Market and Industry Factors in Stock Price Behavior.
{\em Journal of Business} 39(1): 139-190.

\bibitem[Korajczyk and Sadka, 2008]{Q63} Korajczyk, R.A. and Sadka, R. (2008)
Pricing the Commonality Across Alternative Measures of Liquidity.
{\em Journal of Financial Economics} 87(1): 45-72.

\bibitem[Kothari and Shanken, 1997]{Q64} Kothari, S. and Shanken, J. (1997)
Book-to-market, dividend yield and expected market returns: A time series analysis.
{\em Journal of Financial Economics} 44(2): 169-203.

\bibitem[Lakonishok {\em et al}, 1994]{Q65} Lakonishok, J., Shleifer, A. and Vishny, R.W. (1994)
Contrarian Investment, Extrapolation, and Risk.
{\em The Journal of Finance} 49(5): 1541-1578.

\bibitem[Lee and Stefek, 2008]{Q66} Lee, J.-H. and Stefek, D. (2008)
Do Risk Factors Eat Alphas?
{\em The Journal of Portfolio Management} 34(4): 12-24.

\bibitem[Ledoit and Wolf, 2003]{LW1} Ledoit, O. and Wolf, M. (2003)
Improved estimation of the covariance matrix of stock returns with an application to portfolio selection.
{\em Journal of Empirical Finance} 10(5): 603-621.

\bibitem[Ledoit and Wolf, 2004]{LW} Ledoit, O. and Wolf, M. (2004)
Honey, I Shrunk the Sample Covariance Matrix.
{\em The Journal of Portfolio Management} 30(4): 110-119.

\bibitem[Lehmann and Modest, 1988]{Q67} Lehmann, B. and Modest, D. (1988)
The empirical foundations of the arbitrage pricing theory.
{\em Journal of Financial Economics} 21(2): 213-254.

\bibitem[Liew and Vassalou, 2000]{Q68} Liew, J. and Vassalou, M. (2000)
Can Book-to-Market, Size and Momentum be Risk Factors that Predict Economic Growth?
{\em Journal of Financial Economics} 57(2): 221-245.

\bibitem[Lintner, 1965]{Q69} Lintner, J. (1965)
The valuation of risky assets and the selection of risky investments in stock portfolios and capital budgets.
{\em The Review of Economics and Statistics} 47(1): 13-37.

\bibitem[Lo, 2010]{Q70} Lo, A.W. (2010) {\em Hedge Funds: An Analytic Perspective.} Princeton, NJ: Princeton University Press.

\bibitem[Lo and MacKinlay, 1990]{Q71} Lo, A.W. and MacKinlay, A.C. (1990)
Data-snooping biases in tests of financial asset pricing models.
{\em Review of Financial Studies} 3(3): 431-468.

\bibitem[MacKinlay, 1995]{Q72} MacKinlay, A.C. (1995)
Multifactor models do not explain deviations from the CAPM.
{\em Journal of Financial Economics} 38(1): 3-28.

\bibitem[MacQueen, 2003]{Q73} MacQueen, J. (2003)
The structure of multifactor equity risk models.
{\em Journal of Asset Management} 3(4) 313-322.

\bibitem[Markowitz, 1952]{Q74} Markowitz, H.M. (1952)
Portfolio Selection.
{\em Journal of Finance} 7(1): 77-91.

\bibitem[\,1984]{Q75} Markowitz, H.M. (1984)
The Two-Beta Trap.
{\em Journal of Portfolio Management} 11(1): 12-19.

\bibitem[Menchero and Mitra, 2008]{Q76} Menchero, J. and Mitra, I. (2008)
The Structure of Hybrid Factor Models.
{\em Journal of Investment Management} 6(3): 35-47.

\bibitem[Merton, 1973]{Q77} Merton, R. (1973)
An intertemporal capital asset pricing model.
{\em Econometrica} 41(5): 867-887.

\bibitem[Miller, 2006]{Q78} Miller, G. (2006)
Needles, Haystacks, and Hidden Factors.
{\em The Journal of Portfolio Management} 32(2): 25-32.

\bibitem[Motta {\em et al}, 2011]{Q79} Motta, G., Hafner, C. and von Sachs, R. (2011)
Locally stationary factor models: identification and nonparametric estimation.
{\em Econometric Theory} 27(6): 1279-1319.

\bibitem[Mukherjee and Mishra, 2005]{Q80} Mukherjee, D. and Mishra, A.K. (2005)
Multifactor Capital Asset Pricing Model Under Alternative Distributional Specification.
SSRN Working Papers Series, http://ssrn.com/abstract=871398 (December 29, 2005).

\bibitem[Ng {\em et al}, 1992]{Q81} Ng, V., Engle, R.F. and Rothschild, M. (1992)
A multi-dynamic-factor model for stock returns.
{\em Journal of Econometrics} 52(1-2): 245-266.

\bibitem[Pastor and Stambaugh, 2003]{Q82} Pastor, L. and Stambaugh, R.F. (2003)
Liquidity Risk and Expected Stock Returns.
{\em The Journal of Political Economy} 111(3): 642-685.

\bibitem[Rebonato and J\"ackel]{RJ} Rebonato, R. and J{\"a}ckel, P. (1999)
The most general methodology to create a valid correlation matrix for risk management and option pricing purposes.
SSRN Working Papers Series, http://ssrn.com/abstract=1969689 (December 7, 2011).

\bibitem[Roll and Ross, 1980]{Q83} Roll, R. and Ross, S.A. (1980)
An Empirical Investigation of the Arbitrage Pricing Theory.
{\em Journal of Finance} 35(5): 1073-1103.

\bibitem[Rosenberg, 1974]{Q84} Rosenberg, B. (1974)
Extra-Market Components of Covariance in Security Returns.
{\em Journal of Financial and Quantitative Analysis} 9(2): 263-274.

\bibitem[Ross, 1976]{Q85} Ross, S.A. (1976)
The arbitrage theory of capital asset pricing.
{\em Journal of Economic Theory} 13(3): 341-360.

\bibitem[\,1978a]{Q86} Ross, S.A. (1978a)
A Simple Approach to the Valuation of Risky Streams.
{\em Journal of Business} 51(3): 453-475.

\bibitem[\,1978b]{Q87} Ross, S.A. (1978b)
Mutual Fund Separation in Financial Theory -- The Separating Distributions.
{\em Journal of Economic Theory} 17(2): 254-286.

\bibitem[Scholes and Williams, 1977]{Q88} Scholes, M. and Williams, J. (1977)
Estimating Betas from Nonsynchronous Data.
{\em Journal of Financial Economics} 5(3): 309-327.

\bibitem[Schwert, 1990]{Q89} Schwert, G. (1990)
Stock returns and real activity: A century of evidence.
{\em Journal of Finance} 45(4): 1237-1257.

\bibitem[Shanken, 1987]{Q90} Shanken, J. (1987)
Nonsynchronous data and the covariance-factor structure of returns.
{\em Journal of Finance} 42(2): 221-231.

\bibitem[\,1990]{Q91} Shanken, J. (1990)
Intertemporal Asset Pricing: An Empirical Investigation.
{\em Journal of Econometrics} 45(1-2): 99-120.

\bibitem[Shanken and Weinstein, 2006]{Q92} Shanken, J. and Weinstein, M.I. (2006)
Economic Forces and the Stock Market Revisited.
{\em Journal of Empirical Finance} 13(2): 129-144.

\bibitem[Sharpe, 1963]{Q93} Sharpe, W.F. (1963)
A simplified model for portfolio analysis.
{\em Management Science} 9(2): 277-293.

\bibitem[\,1964]{Q94} Sharpe, W. (1964)
Capital asset prices: A theory of market equilibrium under conditions of risk.
{\em Journal of Finance} 19(3): 425-442.

\bibitem[Sharpe, 1994]{Sharpe94} Sharpe, W.F. (1994)
The Sharpe Ratio.
{\em The Journal of Portfolio Management} 21(1): 49-58.

\bibitem[Stock and Watson, 2002a]{Q95} Stock, J.H. and Watson, M.W. (2002a)
Macroeconomic forecasting using diffusion indexes.
{\em Journal of Business and Economic Statistics} 20(2): 147-162.

\bibitem[\,2002b]{Q96} Stock, J.H. and Watson, M.W. (2002b)
Forecasting using principal components from a large number of predictors.
{\em Journal of the American Statistical Association} 97(460): 1167-1179.

\bibitem[Stroyny, 2005]{Q97} Stroyny, A.L. (2005)
Estimating a Combined Linear Factor Model.
In: Knight, J. and Satchell, S.E. (eds.) {\em Linear Factor Models in Finance.}
Oxford: Elevier Butterworth-Heinemann.

\bibitem[Treynor, 1999]{Q98} Treynor, J.L. (1999)
Towards a Theory of Market Value of Risky Assets.
In: Korajczyk, R. (ed.) {\em Asset Pricing and Portfolio Performance: Models, Strategy, and Performance Metrics.}
London: Risk Publications.

\bibitem[Vassalou, 2003]{Q99} Vassalou, M. (2003)
News Related to Future GDP Growth as a Risk Factor in Equity Returns.
{\em Journal of Financial Economics.} 68(1): 47-73.

\bibitem[Whitelaw, 1997]{Q100} Whitelaw, R. (1997)
Time variations and covariations in the expectation and volatility of stock market returns.
{\em Journal of Finance} 49(2): 515-541.

\bibitem[Zangari, 2003]{Q101} Zangari, P. (2003)
Equity factor risk models.
In: Litterman, B. (ed.) {\em Modern Investment Management: An Equilibrium Approach.}
New York, NY: John Wiley \& Sons, Inc., pp. 334-395.

\bibitem[Zhang, 2010]{Q102} Zhang, C. (2010)
A Re-examination of the Causes of Time-varying Stock Return Volatilities.
{\em Journal of Financial and Quantitative Analysis} 45(3): 663-684.

\end{thebibliography}
\end{document}